\documentclass[twocolumn]{aastex6}

\usepackage{mathtools}
\usepackage{epstopdf}
\usepackage{lineno}
\usepackage[varg]{txfonts}
\usepackage{graphicx}
\usepackage{float}
\usepackage[caption=false]{subfig}
\usepackage{amsmath}
\usepackage{commath}
\usepackage[mathscr]{euscript}

\begin{document}

\title{An assessment of and solution to the intensity diffusion error intrinsic to short-characteristic radiative transfer methods}

\author{C. L. Peck\altaffilmark{1,2}, S. Criscuoli\altaffilmark{3}, and M. P. Rast\altaffilmark{2,4}}
\affil{Department of Physics, University of Colorado, Boulder CO 80309, USA\altaffilmark{1}\\
Laboratory for Atmospheric and Space Physics, University of Colorado, Boulder, CO 80303, USA\altaffilmark{2}\\
National Solar Observatory, 3665 Discovery Dr., Boulder, CO 80303, USA\altaffilmark{3}\\
Department of Astrophysical and Planetary Sciences, University of Colorado, Boulder CO 80309, USA\altaffilmark{4}\\
}

\begin{abstract}
Radiative transfer coupled with highly realistic simulations of the solar atmosphere is routinely used to infer the physical properties underlying solar observations. Due to its computational efficiency, the method of short-characteristics is often employed, despite it introducing numerical diffusion as an interpolation artifact. In this paper, we quantify the effect of the numerical diffusion on the spatial resolution of synthesize emergent intensity images, and derive a closed form analytical model of the diffusion error as a function of viewing angle when using linear interpolation. We demonstrate that the image degradation adversely affects the comparison between simulated data and observations, for observations away from disk-center, unless the simulations are computed at much higher intrinsic resolution than the observations.  We also show that the diffusion error is readily avoided by interpolating the simulation solution on a viewing-angle aligned grid prior to computing the radiative transfer.  Doing this will be critical for comparisons with observations using the upcoming large aperture telescopes --- the Daniel K. Inouye Solar Telescope and the European Solar Telescope.   
\end{abstract}


\section{Introduction}
Computational radiative transfer is critical to inferring the physical properties of  
astrophysical objects from observed spectra. Moreover, 
radiation is often a key energy transport mechanism and radiative transfer modeling plays an important role in 
hydrodynamic and magnetohydrodynamic (MHD) simulations of a wide range of phenomena from planetary and stellar atmospheres to accretion disks around compact objects.  Reliable solution of the transfer equation is required to both determine the radiative
heating rate in solution of the energy equation and synthesize spectra that can be compared with observations.

Several techniques have been developed to numerically solve the
radiative transfer equation.  Most often the solution is sought along rays by evaluating the formal solution 
\begin{equation} \label{eq1}
I (\tau_{\nu})=I_{\nu}(0)\, e^{-\tau_{\nu}} + \int\limits_0^{\tau_{\nu}} S_\nu(t_{\nu})\, e^{-(\tau_{\nu} -t_{\nu})}\, dt_{\nu}
\end{equation}
with
\begin{equation} \label{eq2}
\tau_{\nu}=\int_{0}^{L} \kappa_{\rm \nu}\, \rho\, ds\,,
\end{equation}
where $I_\nu$ is the specific intensity at frequency $\nu$, $\tau_{\nu}$ is the optical depth, $\kappa_\nu$ is the frequency specific opacity, and $S_\nu$ is the source function (the ratio of the thermal emissivity to the opacity), which in local thermodynamic equilibrium is taken to be the Planck function (see \citealt{mihalas1984} for details and \citealt{carlsson2008} for a short review of solution methods as applied to a three-dimensional transfer in stellar atmospheres of cool stars).  Moments of the radiation field are then evaluated by numerical quadrature~\citep{carlson1963, lathrop1965, carlson1970} using a limited number of ray directions. 

\begin{figure}[t!]
\centering
	\subfloat{\includegraphics[width=50mm,height=35mm, trim=0mm 0mm 0mm 0mm, clip=true]{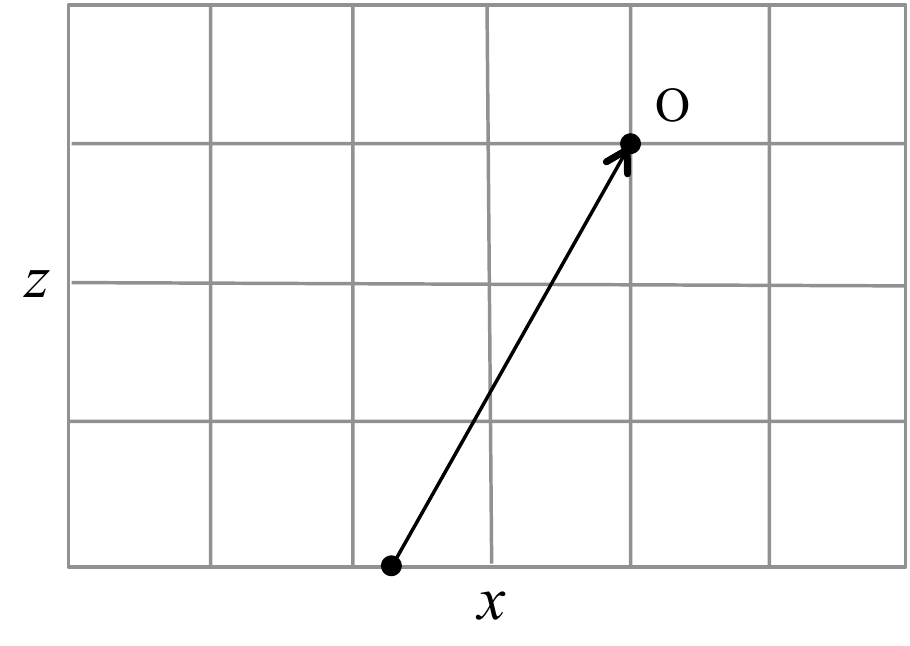}}
	\hspace{5mm}	
	\subfloat{\includegraphics[width=50mm,height=35mm, trim=0mm 0mm 0mm 0mm, clip=true]{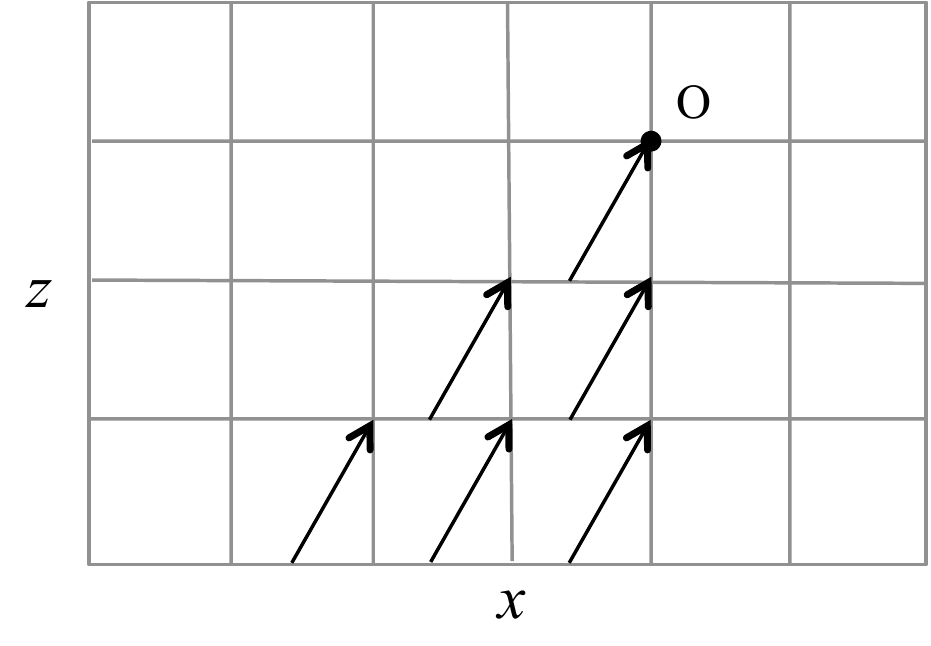}}
	\caption{Ray propagation in the long-characteristic ({\it top}) and short-characteristic ({\it bottom}) methods.}
	\label{SC_interp}
\end{figure}

Integration of the formal solution (Eq.~\ref{eq1}) along rays typically employs one of two strategies: the long-characteristic~\citep{mihalas1978}
or the short-characteristic~\citep{kunasz1988} method. These are illustrated in Fig.~1 
for radiation propagating from the bottom of the domain, where $I_\nu(0)$ is specified, to a point within the domain.  
The long-characteristic method solves
the radiative transfer equation along rays connecting each downwind grid point in the domain (the point for which the specific intensity is needed, position O 
in the top panel of Fig.~\ref{SC_interp}) 
to the last upwind point where the ray originates (bottom of the domain in Fig.~\ref{SC_interp}).  Because the ray does not necessarily intersect the numerical grid except at the upwind point, the long-characteristic method requires interpolation of the plasma properties ($\kappa_\nu$, $\rho$, and $S_\nu$) at ray intersections with the grid rows or columns, depending on ray direction. If estimates of the specific intensity $I_\nu$ are required at
each spatial point in the domain, as is the case in solution of the energy equation or when determining the radiation field in non local-thermodynamic-equilibrium, then, for each ray direction, the number of computations (interpolations) needed for each point on a three-dimensional grid is of order of the number of grid points in a single direction $N$, and the total problem scales as $\mathcal{O}(N^4)$.

The short-characteristic method updates the specific intensity radiation row by row by solving the transfer equation for each grid point along rays starting from interpolated values on the previous row (bottom panel in Fig.~\ref{SC_interp}) or column, depending on ray direction.  
Sweeping the grid in this way reduces the total number of computations needed to update the grid for any ray direction by a factor of $N$ so that the method scales as $\mathcal{O}(N^3)$. 
Because of this reduction in computational cost, the short-characteristic method
is routinely employed in radiative magnetohydrodynamic solvers 
\citep[e.g.][]{vogler2005,hayek2010,gudiksen2011,davis2012} and stand-alone radiative transfer solvers 
\citep[e.g.][]{uitenbroek2001,criscuoli2009,leenaarts2009,stepan2013,ibgui2013,pereira2015,zhu2015}. 

The most significant drawback of the short-characteristic method is that for non vertical or non grid-intersecting ray directions the specific intensity suffers 
diffusion because it is successively interpolated as it is propagated.  In the long-characteristic method only the plasma properties and not the specific intensity itself need to be interpolated to integrate along the ray path.  The diffusive error introduced by the short-characteristics method
has been known since the foundational work of \citet{kunasz1988}, which demonstrated the
occurrence of angular dispersion of the radiation for non grid-aligned propagation directions. The error of the scheme decreases with increasing order of 
interpolation scheme, but at the cost of ringing in regions of steep specific intensity gradients.  This can lead to negative intensity values, which
in turn can be mitigated by monotonic interpolation schemes
 \citep[e.g.][]{auer1994, criscuoli_serena_2007,hayek2010,ibgui2013}, but a consequence of these is the non conservation of the radiative energy~\citep{criscuoli_serena_2007}.  For that reason, and for their computational efficiency, many widely used radiative transfer solves rely on the short-characteristic method with a linear interpolation despite the diffusive error introduced (e.g.; RH ~\cite{uitenbroek2001}, PORTA~\cite{stepan2013}, RH1.5~\cite{pereira2015}, MULTI3D~\citet{leenaarts2009}).  
 
Despite the longstanding knowledge of the intensity diffusion error in the short-characteristic method, little work has been done to fully quantify its effect on simulations.  In simulations the short-characteristic diffusion error propagates through the solution to the radiative heating rate in the energy equation and the solution to the specific intensity --- either solved directly in MHD simulations or post-facto using stand-alone radiative transfer schemes. Such error propagation could lead to angle viewing angle dependent inaccuracies in the numerical solution to the energy equation, as well as spatial diffusion in the emergent intensity. Some work has been done to demonstrate the errors introduced in the radiative heating rate and emergent intensity \citep[e.g.,][]{bruls_1999, kunasz1988}, but only for specific grid geometries. 
Since the simulations are used to infer the physical properties seen in solar observations, intensity diffusion error in the radiative transfer can misguide these inferences. 
Understanding these errors will become increasingly important for measurements made with large aperture telescopes, such as the upcoming NSF's Daniel K. Inouye Solar Telescope (DKIST) \citep{elmore_daniel_2014, tritschler_daniel_2016} and the planned European Solar Telescope (EST) \citep{collados_european_2013, matthews_european_2016}, for which the diffraction limit is nearing the resolution of high-resolution radiative MHD simulations of the solar surface \citep[e.g.,][]{rempel2014, freytag2002, galsgaard1996}.  

In this work we quantify the effective reduction in spatial resolution of the emergent intensity that results when employing a short-characteristic radiative transfer method at inclined viewing angles. We derive a closed form analytical model for the specific intensity of a beam at each point of the grid as a function of beam angle when the beam is initiated as a delta function at the bottom of the computation domain.  The emergent intensity at the top of the domain is, therefore, the effective point spread function of the numerical scheme.  We then validate the model by comparing the intensity obtained by a numerical short-characteristic solution of the radiation emerging from a three-dimensional magnetohydrodynamic simulation snapshot with that predicted by the analytic model. Finally, we demonstrate that the diffusive error is readily avoided by interpolating the simulation atmosphere on a viewing-angle aligned grid prior to computing the radiative transfer solution. Section 2 describes the analytical model. Section 3 assesses the effect of short-characteristic intensity diffusion on spatial resolution of the synthesized emergent intensity, and demonstrates that pre-tilting the simulation atmosphere avoids the diffusion error. Section 4 examines the effect of higher-order interpolation on the intensity diffusion.  

\begin{figure*}[ht!]
\centering
\resizebox{\hsize}{!}{
	\subfloat{\includegraphics[width=75mm,height=60mm, trim=0mm 0mm 0mm 0mm, clip=true]{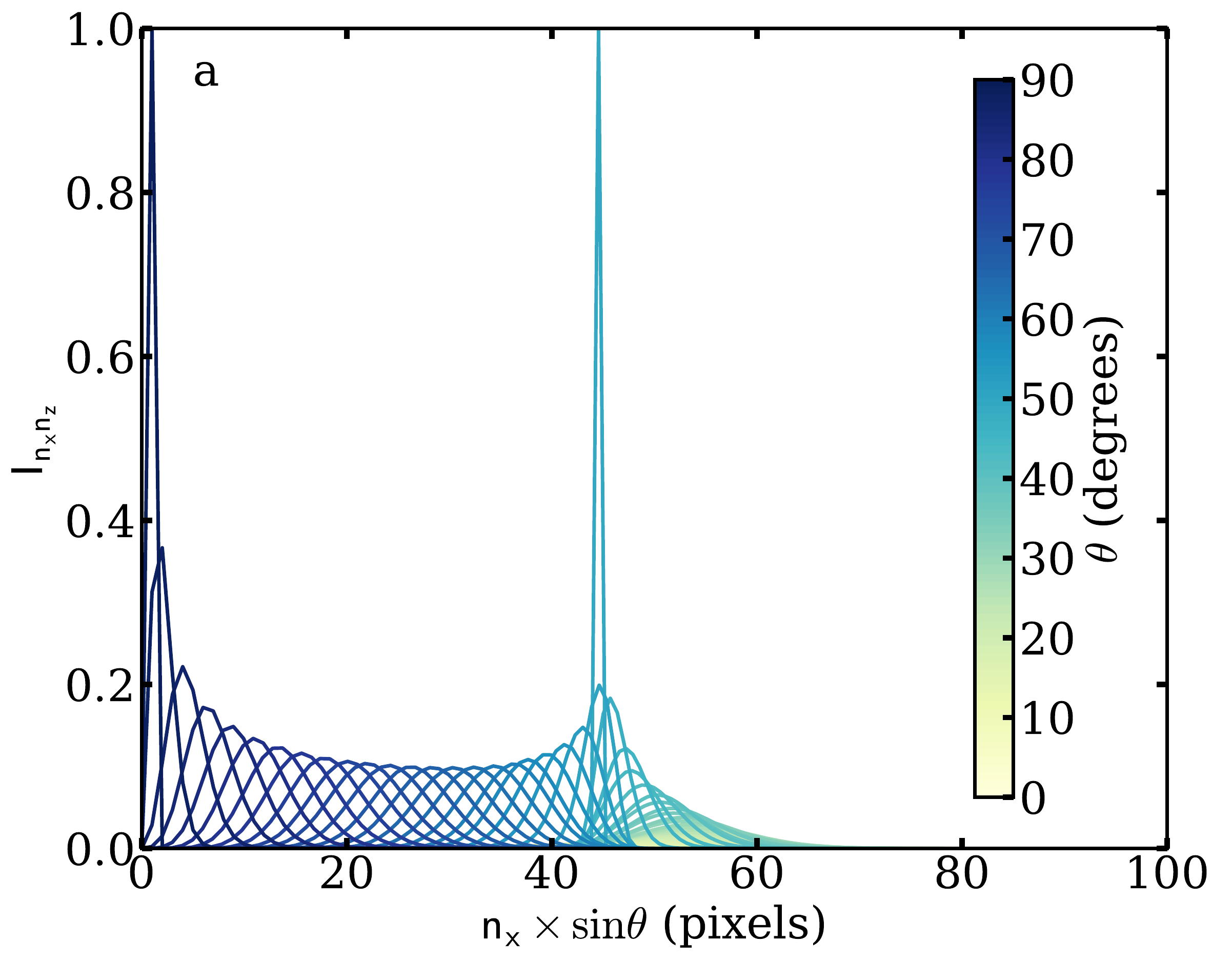}\label{sigma_modela}}
    \hspace{15mm}
    \subfloat{\includegraphics[width=75mm,height=60mm, trim=0mm 0mm 0mm 0mm, clip=true]{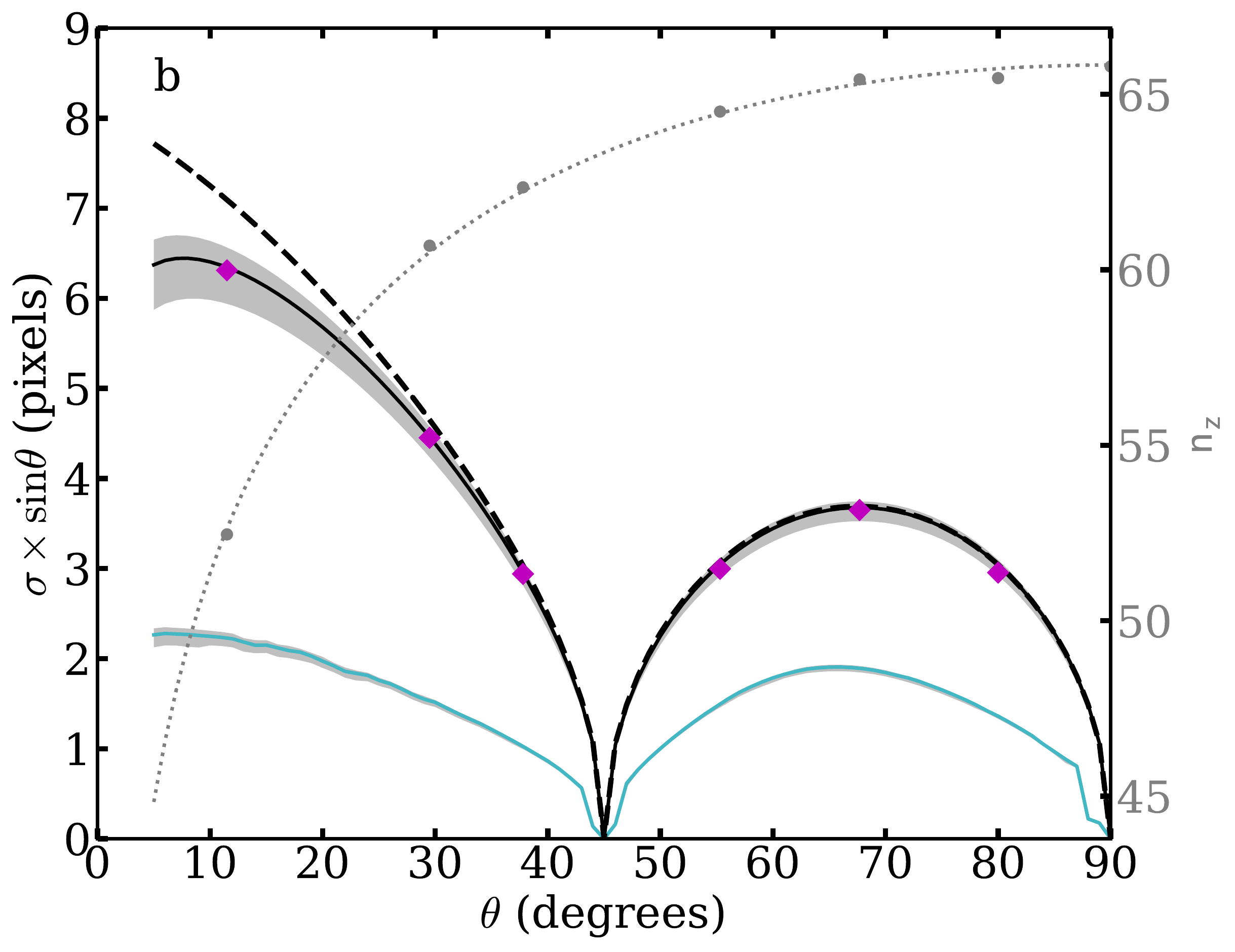}\label{sigma_modelb}}}
	\caption{In ({\it a}), intensity distributions from Eqs.~\ref{eq3} and~\ref{eq4} with $n_z$ = 65. In ({\it b}), the ffective image smearing, computed as the standard deviation of the model (Eq.~\ref{eq5}), for $n_z$ = 65, the depth of the mean vertical ray $\tau=1$ surface in the MHD simulation ({\it thick dashed line}) and for $n_z$ computed using the mean $\tau = 1$ formation height in the MHD model when accounting for viewing angle ({\it solid black line}). The $\tau=1$ formation heights were explicitly computed for viewing angles plotted with \textit{filled grey circles} and determined by a logarithmic fit otherwise (\textit{dotted grey line}). The {\it gray} band indicates the $\sigma$ range for $n_z$ between the $\tau = $ 0.5 and 1.5 depths.  The {\it blue} line and accompanying {\it gray} band indicates the standard deviation for the same $\tau$ surface heights when employing cubic monotonic interpolation in the short characteristic solution.}
	\label{sigma_model}
\end{figure*}

\section{Analytical model of the short-characteristic diffusive error}
\label{Sec:analyticalmodel}
To derive an analytical form for the effective diffusion introduced by the short-characteristic radiative transfer method, we consider a single delta-function point source of radiation at the bottom of a three-dimensional domain and the subsequent propagation of the specific intensity through the domain along inclined ray directions (i.e., the search-beam problem, \citealt{kunasz1988}). To maintain analytical tractability, we examine only linear interpolation on a regular rectangular grid and assume that the beam propagates through vacuum, so that only interpolation of the specific intensity (and not the plasma properties) contributes to diffusion effects. The emergent intensity at the top of the domain is effectively the point spread function of the short-characteristic solution.  Moreover, since, as we will see in more detail, specific intensity interpolation errors are compounded with height, they dominate the error budget even in non-vacuum calculations. 

Depending on the ray propagation direction, the short characteristic method interpolates the specific intensity either on horizontal ($xy$) or vertical ($xz$ or $yz$) planes.  The full derivation of the general three-dimensional solution for arbitrary ray direction is given in Appendix~A.  Here, for simplicity of presentation, we discuss the solution for the special case 
where $\phi$ = 0$^{\circ}$.  This corresponds to ray propagation in the $xz$ plane (as shown in Fig.~\ref{SC_interp}$b$).  
Interpolation then occurs on horizontal (in $x$) or vertical (in $z$) grid lines only, depending on the ray propagation direction $\theta$. Note that $\theta$, the search-beam inclination angle, is defined with respect to the horizontal so that $\theta=90^\circ$ for a vertical propagating ray.  

When $\phi$ = 0$^{\circ}$, the 
three-dimensional solution (Eqs.~\ref{eqna2} and~\ref{eqna6} in Appendix~A) reduces to a point spread function in $x$ only, as no diffusion occurs in the $y$ direction for a ray confined to the $xz$ plane.  For ray angles $45^\circ<\theta < 90^\circ$ (interpolation on horizontal grid lines) the intensity at any grid point can be written as
\begin{align} \label{eq3}
I^{\rm h}_{\rm n_x n_z} &= I_{\rm source} \nonumber \\
 &\times \frac{n_z!}{n_x!(n_z-n_x)!}\bigg(1- \frac{dz}{dx\tan\theta} \bigg)^{n_z-n_x} \bigg(\frac{dz}{dx\tan\theta} \bigg)^{n_x}\,, \nonumber \\
 \end{align}
 where $n_x$ and $n_z$ are the integer number of grid point displacements in the ray direction from the source location.
For ray angles $0^\circ<\theta <45^\circ$ (interpolation on vertical grid lines) the expression becomes 
 \begin{align}\label{eq4}
   I^{\rm v}_{\rm n_x n_z} &=  I_{\rm source} \nonumber \\
  &\times  \frac{n_x-1!}{(n_x-n_z)!(n_z-1)!}\bigg(1-\frac{dx\tan\theta}{dz}\bigg)^{n_x - n_z} \bigg(\frac{dx\tan\theta}{dz}\bigg)^{n_z}\,. \nonumber \\
\end{align}
In these, is the initial point source strength and the specific intensity is the one dimensional point spread (intensity at each grid location $n_x$) caused by the short-characteristic method as a function of beam angle $\theta$ and the number of grid point displacements $n_z$ above the initial source height.  One advantage of writing the intensity distributions in terms of discrete grid point displacements is that they become recognizable as standard and negative binomial distributions in $n_x$ as a function of $n_z$ with standard deviations
\begin{equation} \label{eq5}
\sigma = \sqrt{\ \pm\ n_z  \frac{dz}{dx\tan\theta} \bigg(1 - \frac{dz}{dx\tan\theta}\bigg)}\,,
\end{equation}
where the plus and minus signs apply to the horizontal (Eq.~\ref{eq3}) and vertical (Eq.~\ref{eq4}) grid line interpolation solutions, respectively, and $dz/dx$ is the ratio of the grid scales. The standard deviations represent the spatial smearing or diffusion introduced by the short-characteristic method. 

\begin{figure*}[t!]
\centering
\resizebox{\hsize}{!}{
	\subfloat{\includegraphics[width=20mm,height=57mm, trim=0mm 12mm 00mm 0mm, clip=true]{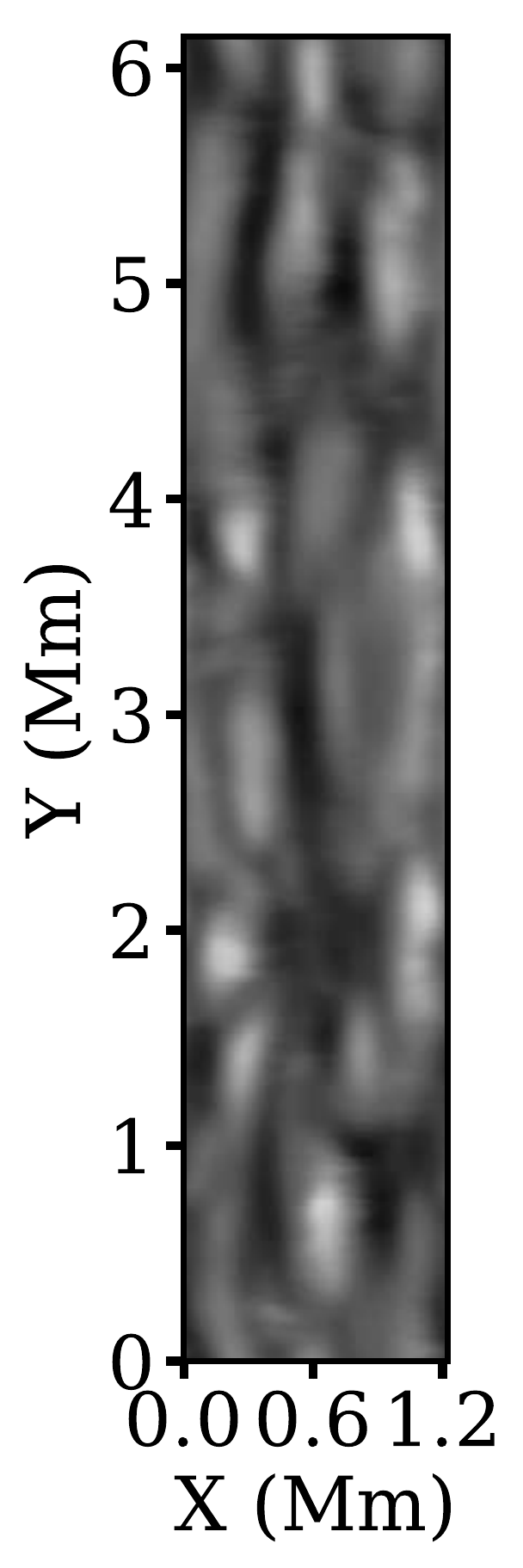}}
    \hspace{2mm}
	\subfloat{\includegraphics[width=30mm,height=57mm, trim=11mm 12mm 00mm 0mm, clip=true]{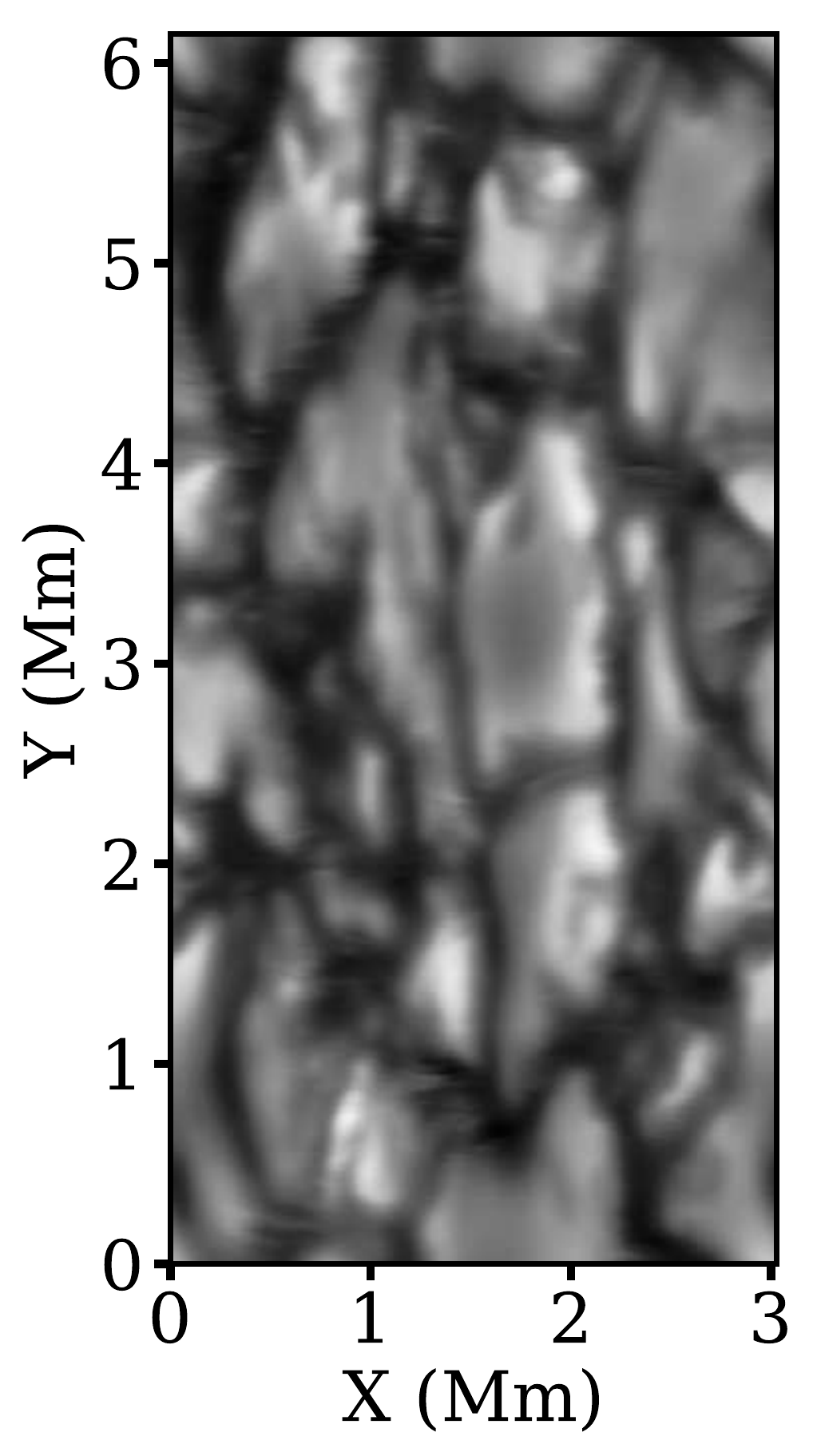}}
    \hspace{2mm}
	\subfloat{\includegraphics[width=45mm,height=57mm, trim=11mm 12mm 0mm 0mm, clip=true]{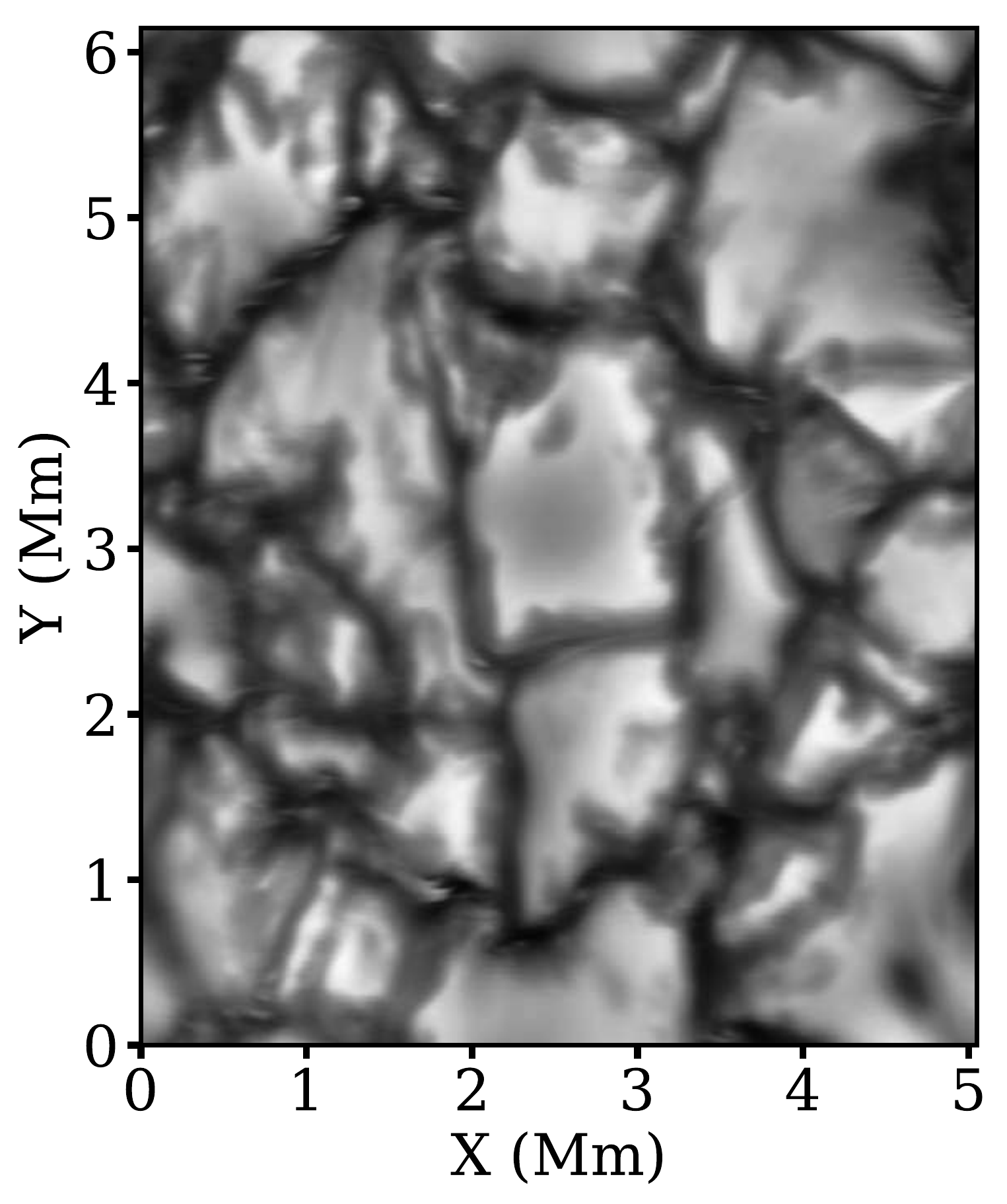}}
    \hspace{2mm}
	\subfloat{\includegraphics[width=53mm,height=57mm, trim=11mm 12mm 00mm 0mm, clip=true]{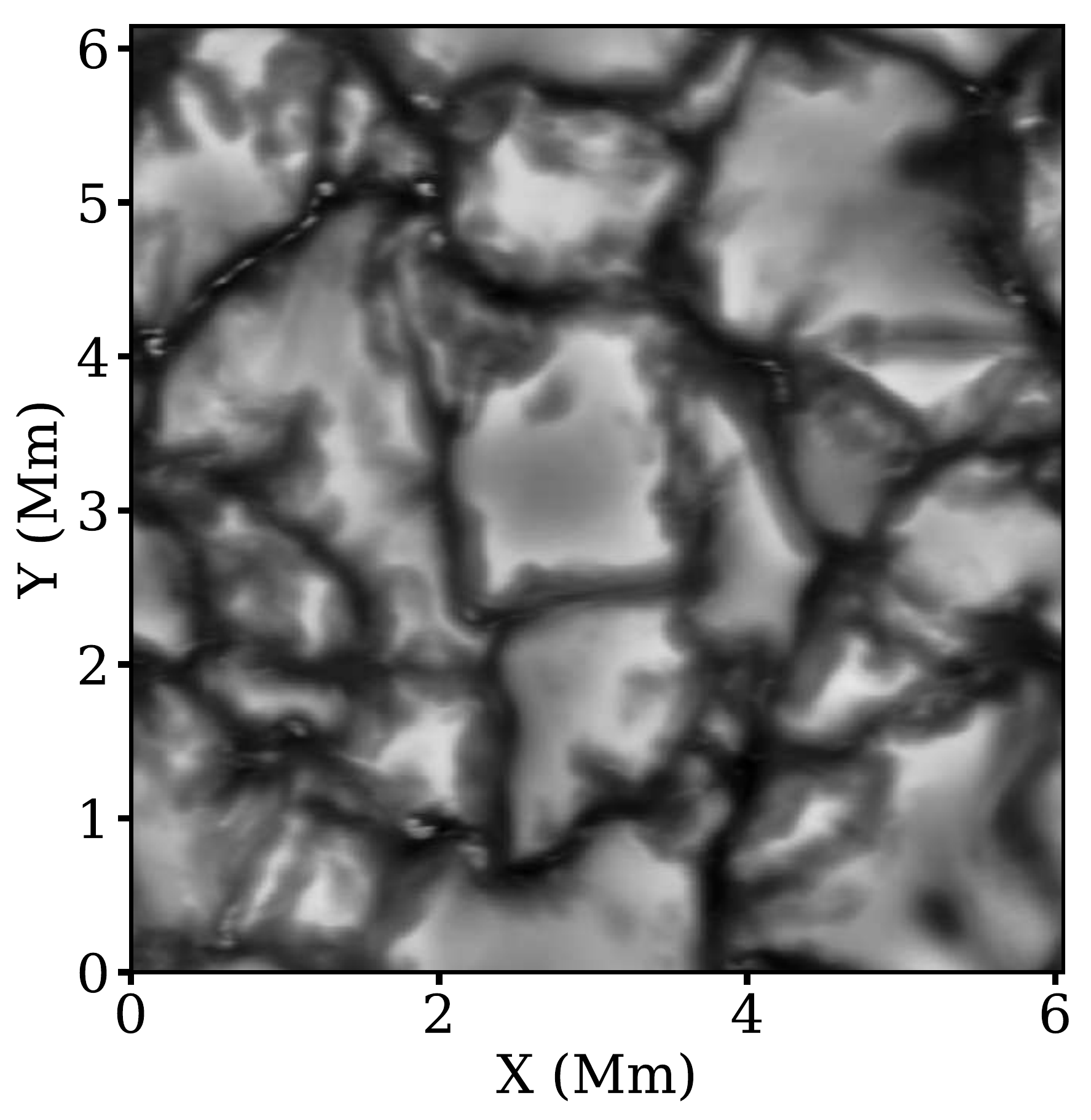}}}\\
\resizebox{\hsize}{!}{
	\subfloat{\includegraphics[width=20mm,height=60mm, trim=0mm 0mm 0mm 0mm, clip=true]{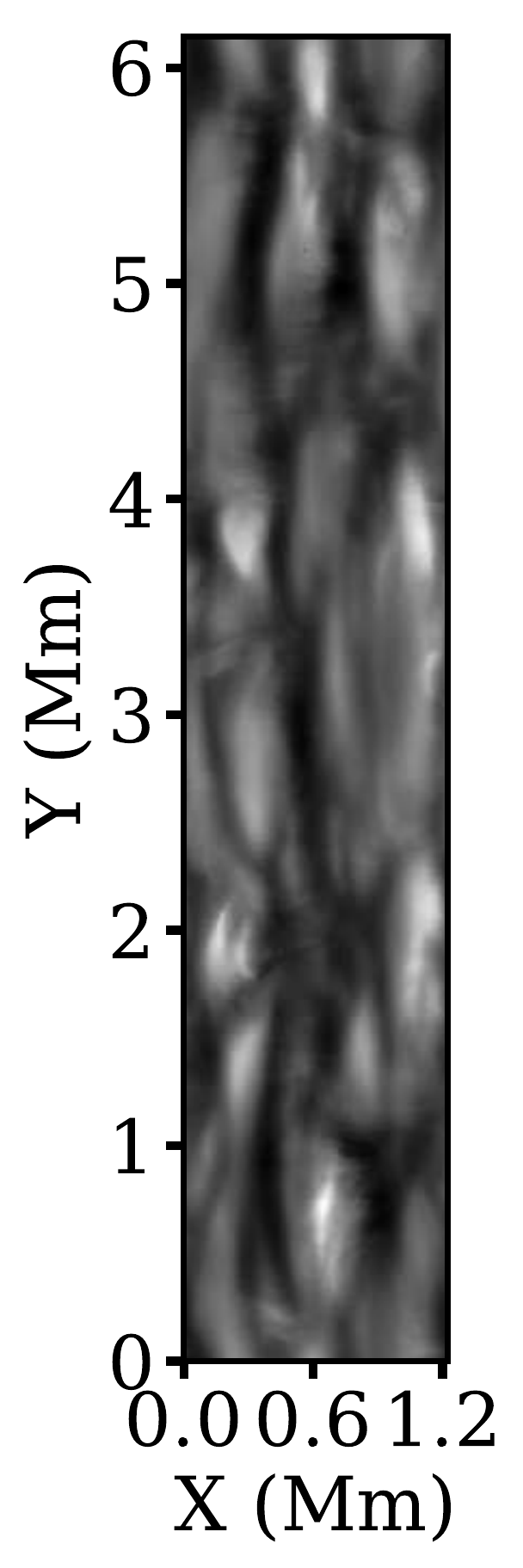}}
    \hspace{2mm}
	\subfloat{\includegraphics[width=30mm,height=60mm, trim=11mm 0mm 0mm 0mm, clip=true]{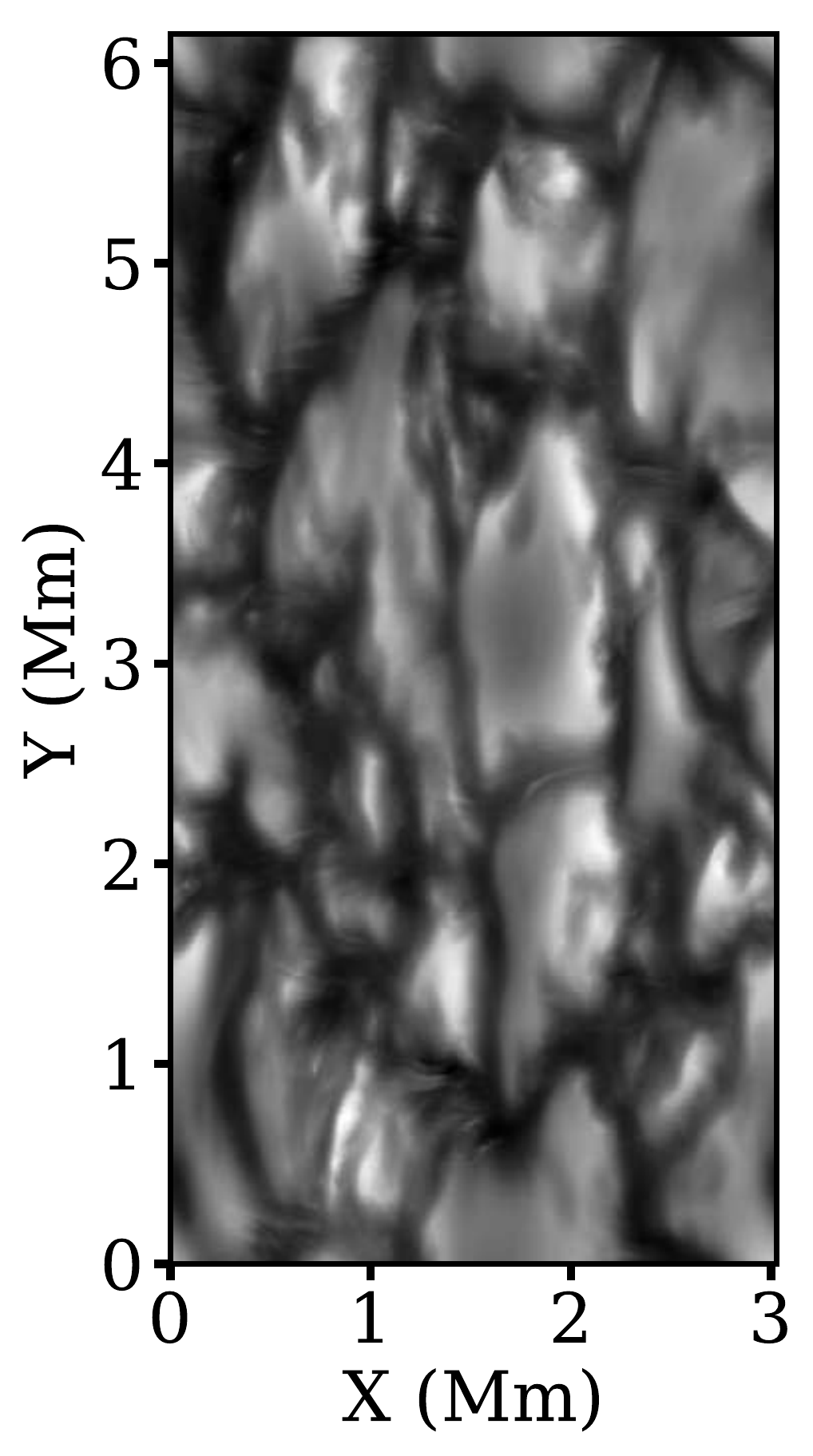}}
    \hspace{2mm}
	\subfloat{\includegraphics[width=45mm,height=60mm, trim=11mm 0mm 0mm 0mm, clip=true]{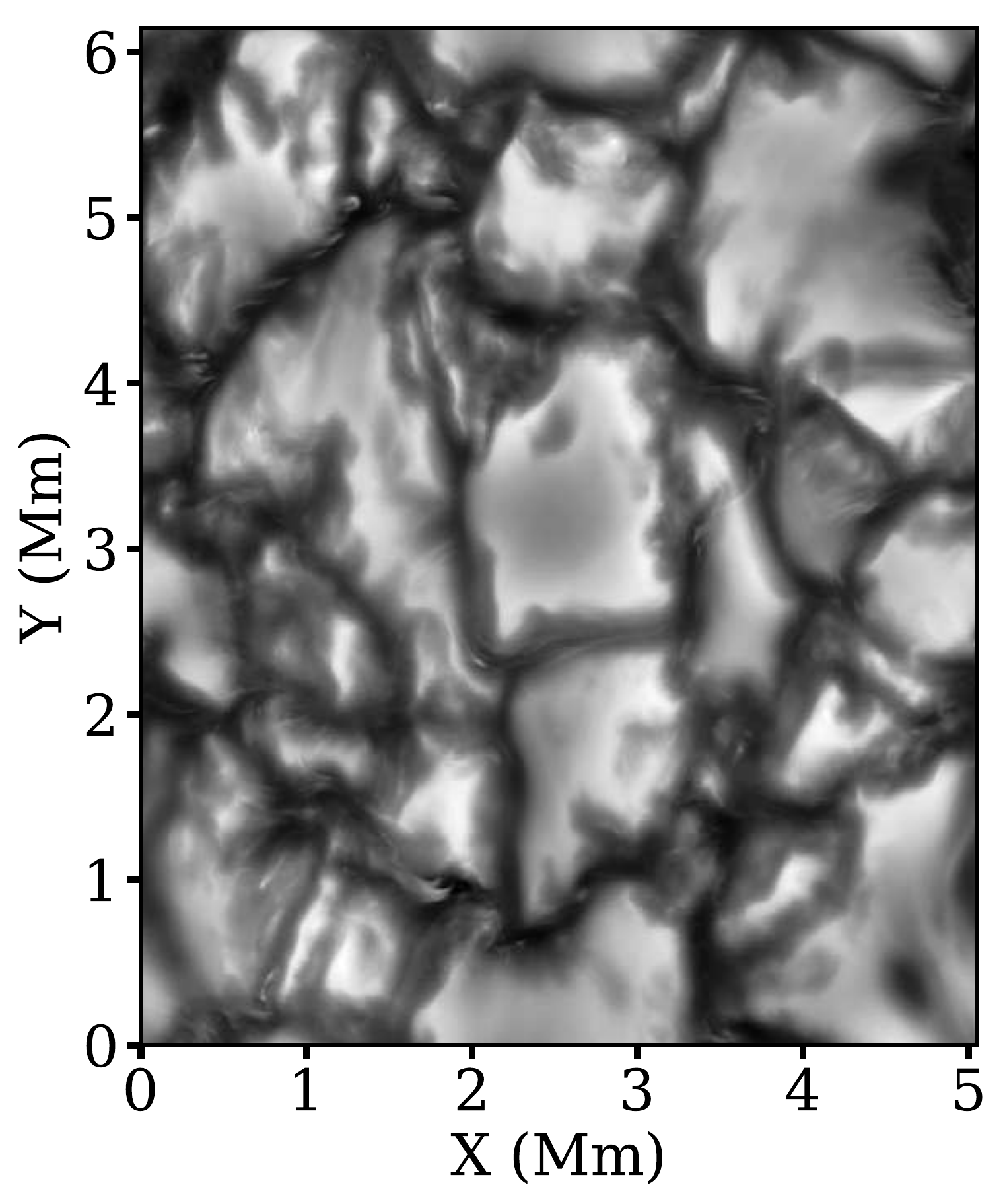}}
    \hspace{2mm}
	\subfloat{\includegraphics[width=53mm,height=60mm, trim=11mm 0mm 0mm 0mm, clip=true]{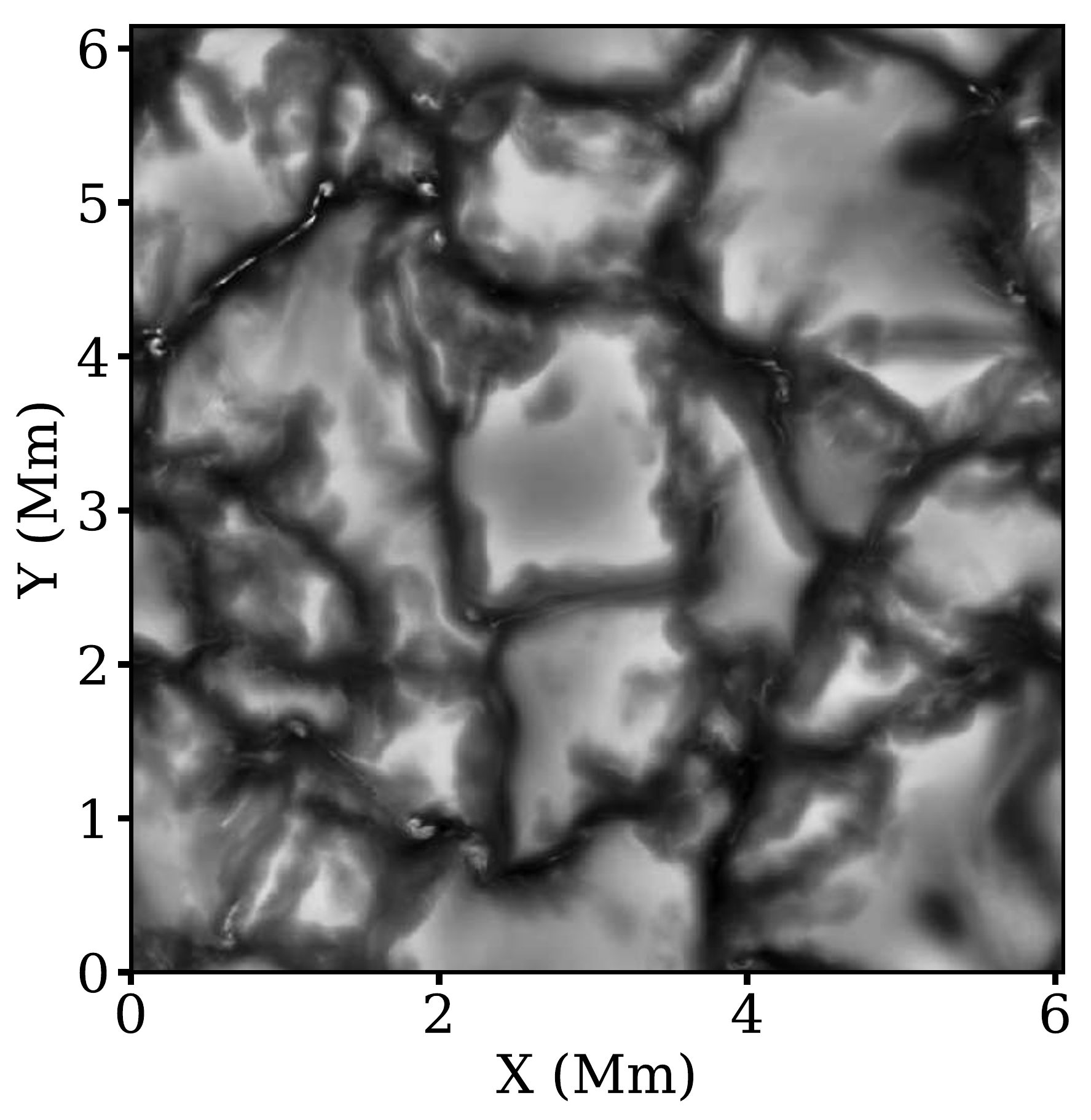}}}
	\caption{Emergent intensity synthesized in the standard short-characteristics method at inclinations values $\mu =$ 0.20, 0.49, 0.82, 0.99 ({\it top row} from {\it left} to {\it right}) . Note the smearing in the direction of the tilt due to the short-characteristic intensity diffusion. {\it Bottom row} shows the emergent intensity for the same inclination angles when synthesized by  on a pre-tilted domain (grid-aligned with the ray propagation direction). Note how the pre-tilted images maintain small scale structures.}
	\label{Images}
\end{figure*}

The intensity distributions and their standard deviations are plotted in Fig.~\ref{sigma_model} as a function of $\theta$ for the case of a square grid ($dx=dz$). 
In these figures $n_x$ and $\sigma$ have been scale by the cosine of the viewing angle to account for foreshortening when the emergent intensity is viewed at inclined angles.  This allows for direct comparison with simulations (Sec.~\ref{sec3}). 
For illustrative purposes, the distributions (Fig.~\ref{sigma_modela}) and standard deviations ({\it thick dashed curve} in \ref{sigma_modelb}) are shown for fixed $n_z = 65$ at multiple ray angles. 
The short-characteristic ray direction is grid aligned at 90 and 45 degrees, so the specific intensity distribution collapses to a delta function at those angles (no diffusive error).  This is true of zero degrees as well, but that case is pathological because a strictly horizontal ray never reaches the upper boundary.  Moreover, vertical grid line interpolation ($0^\circ<\theta <45^\circ$) requires an increasing number of interpolations (to cross an equal number of horizontal grid lines, $n_z$) with decreasing inclination angle so the diffusive error increases monotonically as $\theta \rightarrow 0$.  For horizontal grid interpolation ($45^\circ < \theta < 90^\circ$), the maximum diffusive error $\sigma$ occurs at $67.5^\circ$, half-way between the grid aligned directions.

The specific intensity distributions given by the analytic solutions of Eqs.~{\ref{eq3}} and~\ref{eq4} (more generally by Eqs.~\ref{eqna2} and~\ref{eqna6} in Appendix~A) and plotted in Fig.~\ref{sigma_modela} represent the point spread functions of the short-characteristic solution for any ray propagation angle $\theta$ through a vacuum domain.  The width of the distribution $\sigma$ (Eq.~\ref{eq5} and Fig.~\ref{sigma_modelb} {\it thick dashed curve}) captures the effective image smearing introduced when computing the emergent intensity.  It depends explicitly on $n_z$, the number of interpolations above the source point.  Since the model assumes ray propagation through a vacuum, the analytic solution captures image degradation in a realistic three-dimensional solution due to radiation propagation above the optically thick to optically thin transition.  
Below the $\tau=1$ transition the medium is optically thick and the diffusive error of the short-characteristic method, while it may contribute to the accuracy of radiative heating and cooling terms computed in the radiative magneto hydrodynamic model, does not contribute to degradation of the emergent intensity.  
Thus the diffusive error in any direction depends on the depth within the domain from which the radiation escapes.  
Since the optical depth surfaces are not aligned with the numerical grid, that depth depends not just on the simulation solution, but on the viewing angle and the wavelength of the radiation.  These  must be accounted for when evaluating the short-characteristic error in the emergent intensity from a simulation solution. 

\section{Diffusive error in the emergent intensity from 3D MHD simulations}
\label{sec3}
To assess the amount of image degradation in synthesized emergent intensity images computed from three-dimensional magnetohydrodynamic simulations, we examine a snapshot from a MURaM~\citep{vogler2005} simulation of a $6\times6\times1.2$~Mm region of solar granulation including magnetic field generation by small-scale dynamo processes~\citep{rempel2014}.   The solution was computed at 16~km resolution (8~km grid spacing) in both the vertical and horizontal directions, the latter comparable to the highest resolution images that will be forthcoming with the DKIST. It thus provides a good test bed for future comparisons between simulations and observations.  The synthesized emergent intensity was computed using the RH radiative transfer solver~\citep{uitenbroek2001} for viewing angles between $\mu$ of 0.2 and 1 ($\mu  = \sin\theta$, with $\theta$ as defined previously) and at $\lambda=500 {\rm nm}$.
For simplicity of presentation, as above, we confine the viewing angle to the $xz$ plane ($\phi = 0^{\circ}$), though the more general solution is presented in Appendix~A. 

We perform the radiative transfer in two ways.
In the first, we solve the transfer equation using the standard short-characteristics scheme with the specific intensity linearly interpolated along the inclined rays. The emergent intensities at  $\lambda = $ 500~nm  and viewing angles $\mu=0.20, 0.49, 0.82, 1.00$ are shown in the top row of Fig.~\ref{Images}.  The effect of numerical diffusion in the direction of  inclination is visually apparent in the images.  In the second, we compute the radiative transfer on the same atmosphere by pre-tilting the atmosphere to the required viewing angle, interpolating the atmospheric properties along the viewing angles, and solving the transfer along the now grid-aligned rays. In this way we avoid the specific intensity interpolation error --- though the plasma properties are still interpolated, as required by all radiative transfer solvers. The important point is that the pre-tilting procedure eliminates the compounded specific intensity interpolation error without adding additional sources of error.  The bottom row of Fig.~\ref{Images} shows the emergent intensity when computed using the pre-tilted atmospheres  to the same viewing angles.  It is clear that the pre-tilted images retain better spatial resolution than images obtained with the standard short-characteristic method.

Determining the image smearing using the model of Sec. 2 requires that $n_z$ be specified in Eqs.~\ref{eq3} and ~\ref{eq4}.  As discussed previously, $n_z$ is the number of interpolations suffered between the radiation escape and the observer's location (approximately the number of grid levels along the ray between the $\tau=1$ surface to the top of the domain $n_z$).  We measure the depth of the mean $\tau=1$ surface in the simulation at select viewing angles and plot them as $N_z-n_z(\tau=1)$ using filled circles in Fig.~\ref{sigma_modelb}, where $N_z = 130$ for this simulation. 
The measured values of $\tau = 1$ range from around 77 to 64 grid levels for lines of sight of $10^\circ$ and vertical, respectively, corresponding to $N_z-n_z(\tau=1)$ values of around 53 and 66, respectively. 
For viewing angles between these we use a logarithmic fit to those points ({\it dotted curve} in Fig. 2b), and use the fit to plot the $\sigma$ from Eqs.~\ref{eq3} and ~\ref{eq4} as {\it continuous black} curve in Fig.~\ref{sigma_modelb}. 
A comparison of the $\sigma$ values computed in this 3D MHD snapshot, for which $n_z$ is a function of the inclination angles, with those found for the case of a ray propagating in vacuum discussed in Sec. \ref{Sec:analyticalmodel}, for which  $n_z$  is not a function of $\theta$ ({\it thick dashed line} in Fig.~\ref{sigma_modelb}), shows clearly the dependence of the amount of smearing on the formation height of radiation. In these particular examples, for which the values of $n_z$ coincide for vertical directions, the amounts of smearing  differ significantly  for angles smaller than $\theta \lesssim 30^{\circ}$ .
  
\begin{figure}[t!]
\centering
	\subfloat{\includegraphics[width=80mm,height=65mm, trim=0mm 0mm 0mm 0mm, clip=true]{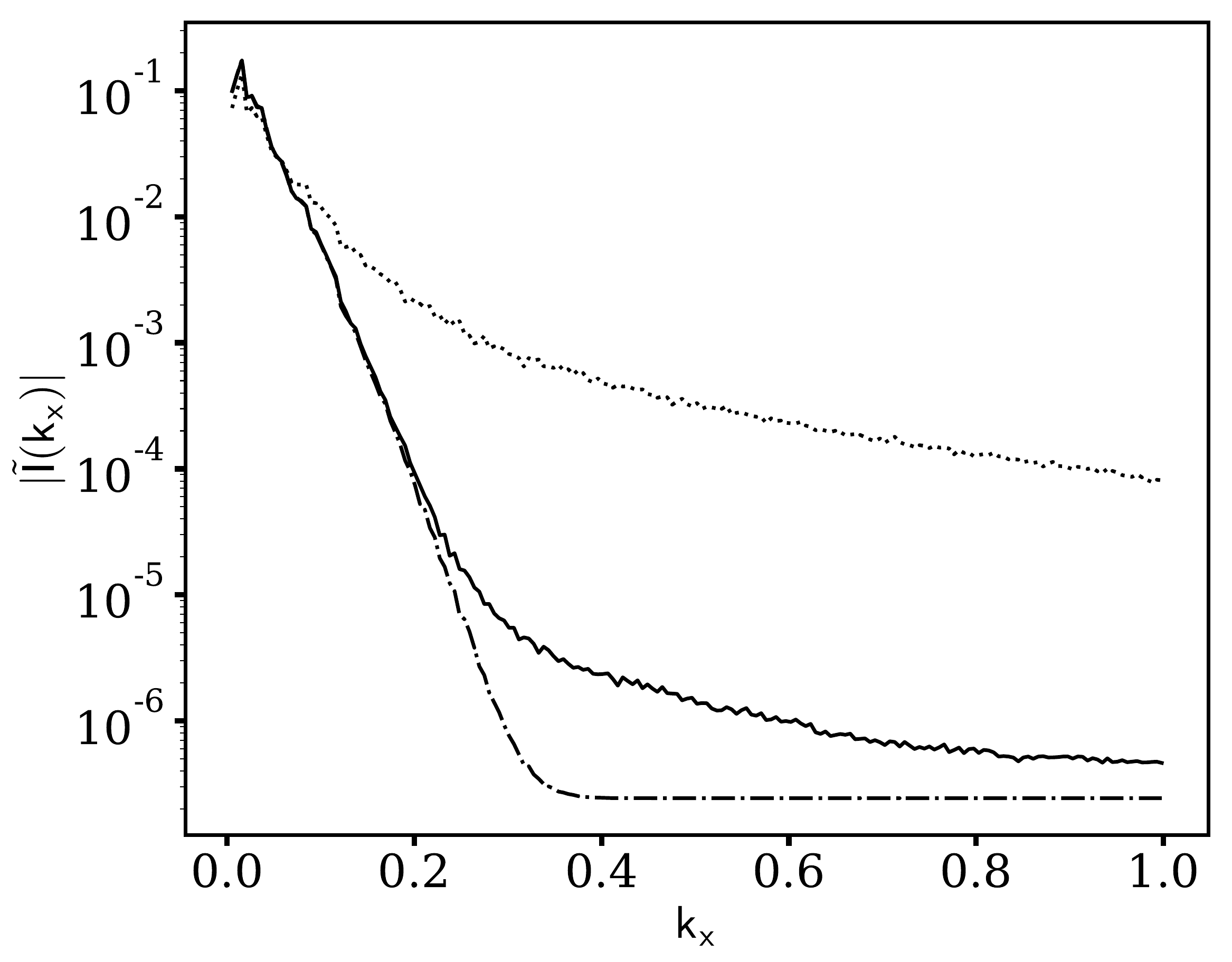}}
	\caption{Average (over image rows) intensity amplitude spectrum for $\mu = 0.49$ after employing the standard short-characteristic method ({\it solid line}) and pre-tilted short-characteristic method ({\it dotted line}).  {\it Dot-dashed line} shows pre-tilted spectrum convolved with a binomial convolution kernel with width that yields the best-fit to the standard spectrum.   Spectra are normalized to share the same integrated amplitude.}
	\label{fft_fits}
\end{figure}

To measure the spatial resolution degradation induced by the standard short-characteristic method using the images themselves, and thus demonstrate that it agrees with the model of Sec. 2 and can be eliminated by pre-tilting and interpolating before computing the transfer, we compare the power spectra of the two image sets.
Since the error occurs only in the inclined direction of the viewing angle prescribed, we compute the average 1D spatial power spectra along the direction of the tilt ($x$-dimension). We then convolve the one-dimensional spectra of the pre-tilted solution with the analytic solution (Eqs.~\ref{eq3} or ~\ref{eq4}) and use a least-squares measure of the difference between the two spectra to determine the standard deviation of the binomial convolution kernel.  An example of the one-dimensional spectra used in this procedure is plotted in Fig.~\ref{fft_fits}, and the estimates of the standard deviation $\sigma$ of the binomial convolution kernel are shown in Fig.~\ref{sigma_modelb} with {\it pink diamonds}. 
The error in those measures, estimated as the uncertainty in the fit between the spectra, is smaller than the symbol size, and the values agree with the image degradation caused by inclined ray interpolation in the short-characteristic method.  Moreover, the agreement demonstrates that the degradation can be avoided by pre-tilting the solution before computing the short-characteristic radiative transfer.

\section{Extension to Higher-Order Interpolation}
\begin{figure}[t!]
\centering
	\subfloat{\includegraphics[width=80mm,height=65mm, trim=0mm 0mm 0mm 0mm, clip=true]{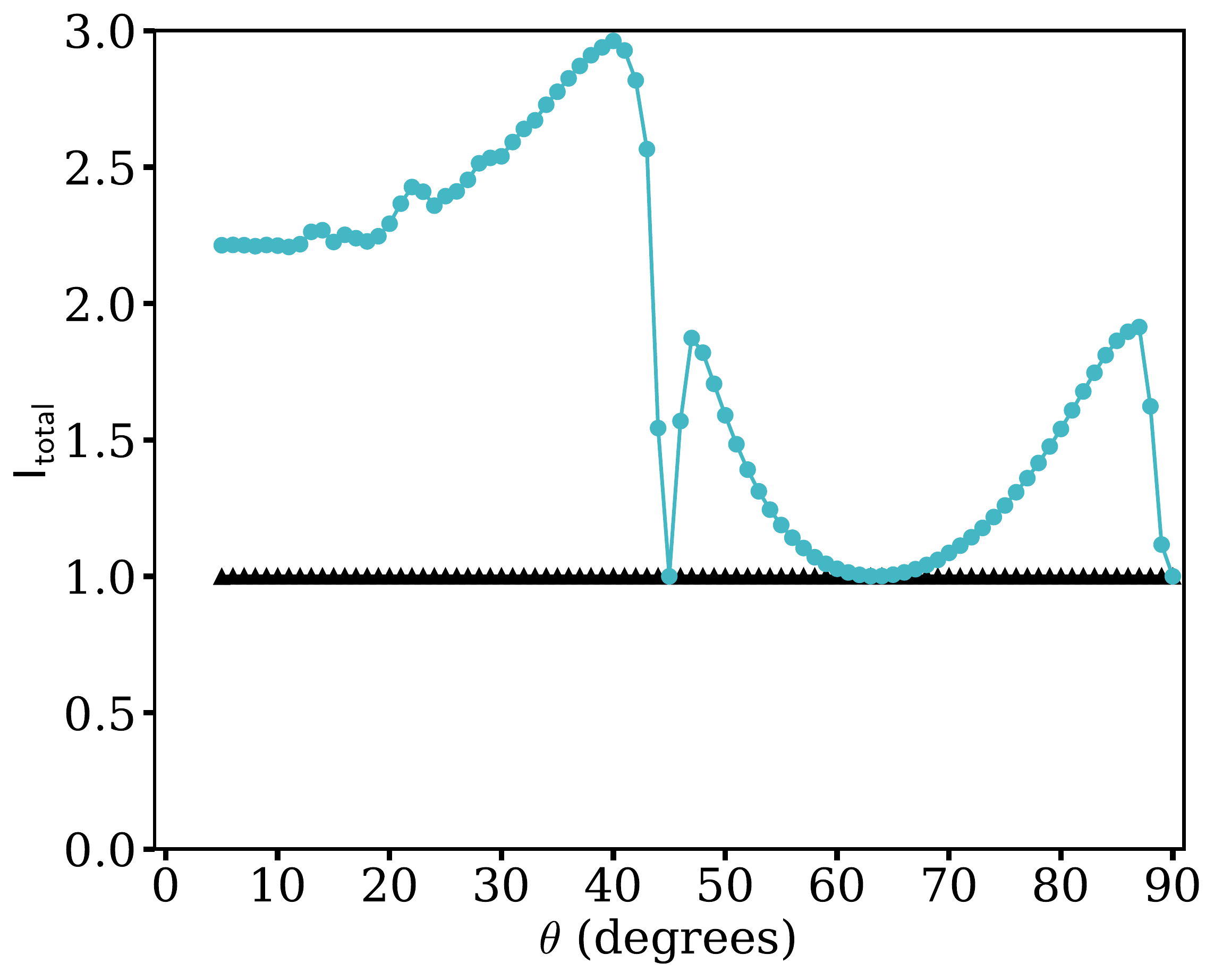}}
	\caption{Normalized integrated intensity for linear ({\it black triangles}) and cubic monotonic ({\it blue circles}) interpolation. Note how the integrated intensity is conserved using linear interpolation, but not using cubic monotonic interpolation.}
	\label{sigma_model2}
\end{figure}

As discussed in Sec.~1, the interpolation error introduced by the short-characteristics method can be reduced by employing a  higher-order monotonic interpolation scheme~\citep[e.g.,][]{kunasz1988, hayek2010}.  To assess the effectiveness of this approach, we revisit the search beam problem, this time employing cubic monotonic interpolation~\citep{fritsch_1980} in the short characteristic solution.  Unlike for linear interpolation solution, we were unable to derive an analytical solution for the intensity on the grid, and so instead propagated the search beam through the vacuum domain numerically.  For direct comparison with the linear interpolation results, the short characteristic solution was iterated $n_z$ times based on the depth of the $\tau$ surfaces measured as a function of angle in the previous section. The width of the resulting intensity distribution was then determined by a skew-normal fit,  and the resulting values are plotted in blue in Fig.~\ref{sigma_modelb}.  Numerical diffusion is significantly reduced (by a factor of up to about three) by employing the higher order interpolation scheme, but because of the monotonicity constraint needed to avoid negative specific intensity values, energy is not conserved.  The integrated intensity at the top of the domain (normalized by the source value) as a function of beam angle is plotted for both linear and cubic monotonic interpolation in Fig.~\ref{sigma_model2}.  While the total intensity is conserved by linear interpolation for all inclination angles (black triangles), cubic monotonic interpolation (blue circles) conserves intensity only at $45^{\circ}$,  $90^{\circ}$, and near $62^{\circ}$.  For all other beam inclinations, the total emergent intensity is overestimated.

\section{Discussion and Conclusion}

We have quantified the effect of interpolation errors inherent in the short-characteristic radiative transfer method, deriving an analytical model of the diffusion of a point source through a grid at arbitrary viewing angle and comparing that with the emergent intensity spatial resolution 
degradation of a full three-dimensional radiative magneto hydrodynamic solution. We have shown that,
because of the compounding nature of the error, intensity interpolation accounts for most, if not all, of the reduction in resolution introduced by the short-characteristic method.  Interpolation of the plasma properties along the ray path, common to both short and long characteristic methods, likely contributes negligibly to resolution reduction. While higher-order interpolation schemes reduce the diffusion error inherent in the short characteristic method, they do not conserve energy.  We have demonstrated that both of these problems can be circumvented by pre-tilting the computational domain to a ray-aligned grid and performing the short-characteristic radiative transfer vertically through this domain. This requires only that the same interpolation of the atmospheric properties along the ray directions occurs prior to the ray propagation, rather than after it. 

Without such pre-tilting, the short-characteristic scheme introduces errors that challenge comparisons between numerical solutions and observations. 
The errors are a strong function of viewing angle, the image and simulation resolutions, and the formation height of the wavelength of interest.  
For simulations and observations with matching native resolution, {\it any} non grid-aligned viewing angle yields a fundamental mismatch in the effective spatial resolution. 
For example, the simulation solutions used in this work has a native resolution of 16~km, which is reduced to 80~km when viewed at $30^{\circ}$. 
The change is equivalent to the difference  in spatial resolution of observations obtained in the visible with telescope apertures of 4 m (e.g. DKIST) and  $\sim$ 0.8 m (e.g. the NSF's Dunn Solar Telescope). 
We note that since the point spread function of the short-characteristic diffusion error is now known analytically (for linear interpolation), the emergent intensity could simply be de-convolved to obtain a fully resolved image for comparison with observations. This is true in the case of local thermodynamic equilibrium where the specific intensity along a ray depends 
only on the local plasma properties, but it is not true for non-local thermodynamic equilibrium for which the radiation field must be iteratively solved --- propagating error throughout the solution to the radiation field.
Employing higher-order interpolation schemes yields an improvement (to 28~km resolution in the example above for the particular monotonic cubic interpolation employed) at the expense of specific intensity non-conservation.  This non-conservation of energy does not affect the emergent intensity image contrast at any given viewing angle, 
but it does affect the relative intensity at differing angles, again likely posing difficulties for non-local thermodynamic equilibrium solutions.

Similarly, since some radiative magneto hydrodynamic solvers use short-characteristic radiative transfer in combination with discrete angle-weighted quadrature schemes to evaluate the divergence of the radiative flux in the solution of the energy equation, angle dependent specific intensity errors may introduce computational artifacts in the radiative magneto hydrodynamic solutions themselves. The effects of numerical diffusion on the moments of the radiative transfer equation were discussed in \citet{bruls_1999} for the case of specific triangular grids. Similarly, angle dependent artifacts have been noted in solutions for the photon density in cosmological solutions for optically thin radiative transfer~\citet{finlator2009}.  A detailed analysis of how short-characteristic error propagates through the quadrature schemes employed by magneto hydrodynamic codes in the optically thick to optically thin transition of solar photosphere is warranted, and the subject of future work. 

Finally, we reiterate that the diffusive errors introduced by the short-characteristic method can be avoided by interpolating the atmosphere onto a ray-aligned grid before computing the transfer. Since the total number of interpolations incurred by pre-tilting is less than that for the standard short-characteristic method (only the plasma properties not the specific intensity values must be interpolated if pre-tilting is employed) the pre-tilting method likely introduces no additional computation time to the radiative transfer solution, though when needed for the quadrature calculation of the flux divergence, interpolation of the specific intensity back to the original grid is also required. This step, however, does not compound error as in the standard short-characteristic method.  Other solutions to the numerical diffusion problem have been explored.  Long-characteristic solvers are routinely employed (e.g.; Bifrost ~\citep{gudiksen2011}, STAGGER \citep{galsgaard1996}, StellarBox \citep{wray_2015}), and these fundamentally avoid the intensity diffusion error. Hybrid radiative transfer schemes using adaptive mesh refinement have also been developed \citep{Rijkhorst_hybrid_2006} and help mitigate the error.  Which approach proves most accurate and computationally efficient is still an open question, but pre-tilting is simple, highly effective, and can perhaps be seamlessly integrated into existing magneto hydrodynamic solvers. 
 
\acknowledgements
This material is based upon work supported by the National Science Foundation Graduate Research Fellowship Program under grant No. DGE 1144083 and the National Science Foundation under grant No. 1616538. The National Solar Observatory is operated by the Association of Universities for Research in Astronomy under a cooperative agreement with the National Science Foundation. This work utilized the Janus supercomputer, which is supported by the National Science Foundation (award number CNS-0821794), the University of Colorado Boulder, the University of Colorado Denver, and the National Center for Atmospheric Research. Janus is operated by the University of Colorado Boulder. The authors would like to thank Matthias Rempel for providing the MHD snapshots and Han Uitenbroek for useful feedback throughout the project. The authors are grateful for the useful discussions at the International Team 335 ``Towards New Models of Solar Spectral Irradiance Based on 3D MHD Simulations", and would like to thank the International Space Science Institute in Bern for hosting the team. 
\clearpage

\appendix
\renewcommand{\thesubsection}{A.\arabic{subsection}}
\renewcommand{\theequation}{A-\arabic{equation}}
\setcounter{equation}{0}  

\begin{figure*}[t!]
\centering
	\subfloat{\includegraphics[width=80mm,height=50mm, trim=0mm 0mm 0mm 0mm, clip=true]{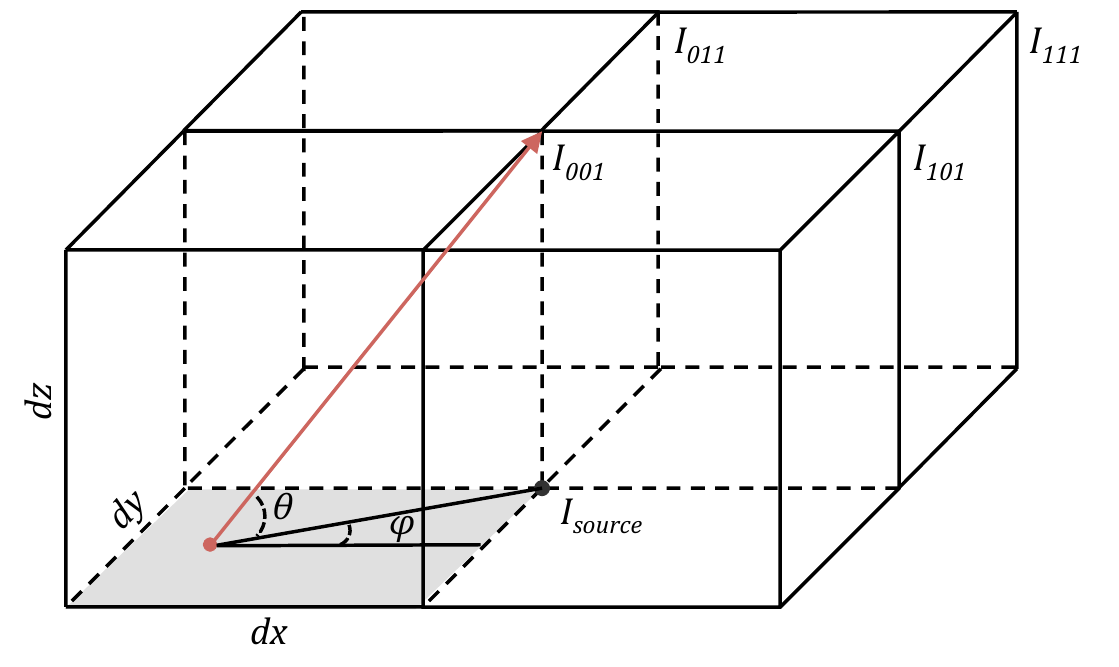}}
    \subfloat{\includegraphics[width=80mm,height=50mm, trim=0mm 0mm 0mm 0mm, clip=true]{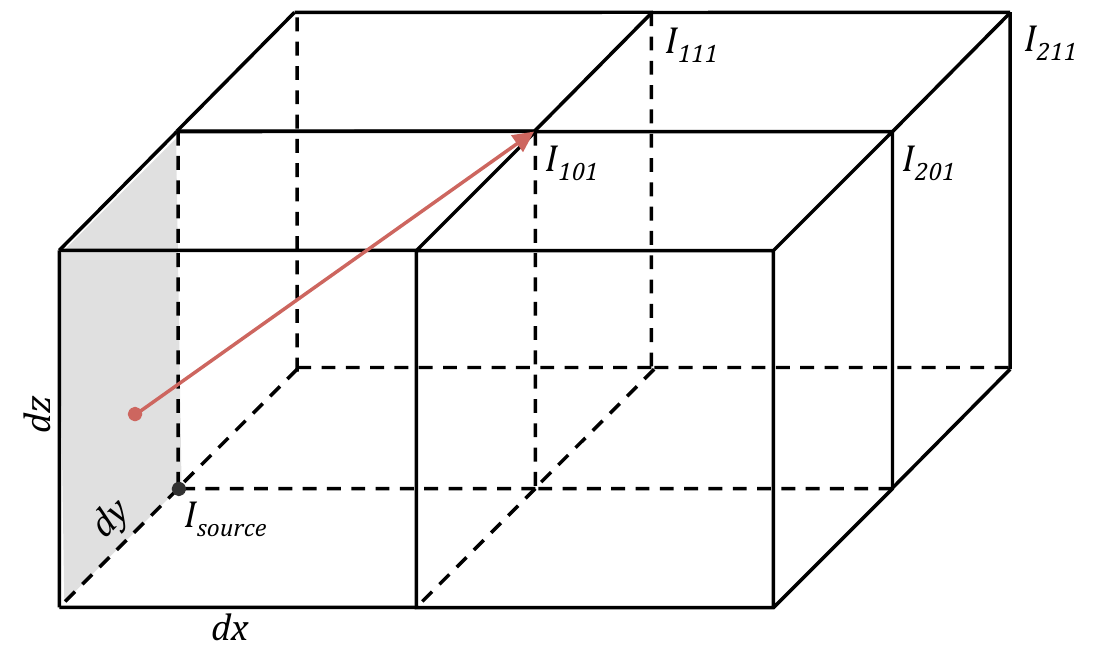}}
	\caption{Short characteristics intensity interpolation in a 3D domain. Left: Interpolation on the {\it xy} plane. Right: Interpolation on th {\it yz} plane. Interpolation on th {\it xz} plane is not shown, but follows that on {\it yz}.  Angles defined as shown in the left image. }
	\label{SC_diffusion}
\end{figure*}

The analytical form of the diffusion error intrinsic to the short-characteristic method is derived here assuming a single delta-function point source of intensity at the bottom of the domain. The ray propagates through vacuum so there is no need to interpolate the plasma properties. To maintain analytical tractability, we assume linear interpolation and a \textit{regular} spatial grid with sampling \textit{dx} by \textit{dy} and \textit{dz} in the horizontal and vertical directions respectively.
In the short-characteristic method, rays are propagated differently depending on whether the upwind intensity is interpolated in a vertical or horizontal plane. We present the two cases separately.  For the ray orientations shown in Fig.~\ref{SC_diffusion}, the intensity for both cases is updated from left to right and then from front to back. The solution is applicable to other ray orientations as the indices used represent the number of grid points away from the $I_{\rm source}$ location (e.g., $I_{\rm 000}$ represents the source location and $I_{\rm n_xn_yn_z}$ represents the point $n_x$, $n_y$, and $n_z$ grid steps from $I_{\rm source}$ in the x,y, and z directions). The domain is taken to have sufficient horizontal extent such that the fully dispersed beam intersects the upper domain boundary before exiting the sides.  If this is not the case, horizontal periodicity must be imposed on the solution.

\subsection{Solution for ray directions for which interpolation occurs on horizontal planes} 

The ray propagation for this subset of angles is shown on the left of  Fig.~\ref{SC_diffusion}. In the 3D geometry, the boundary condition or the previously interpolated values on the upwind $xy$-plane gridpoints provide four intensity values that are linearly interpolated for the intensity at the downwind grid point as: 
\begin{align}\label{eqna1}
	I_{\rm 001} &= I_{\rm source} \bigg(1-\frac{dz}{dx}\frac{\cos\phi}{\tan\theta}\bigg) \bigg(1-\frac{dz}{dy}\frac{\sin\phi}{\tan\theta}\bigg) \nonumber\,, \\ 
	I_{\rm 101} &= I_{\rm source}  \bigg(\frac{dz}{dx}\frac{\cos\phi}{\tan\theta}\bigg) \bigg(1-\frac{dz}{dy}\frac{\sin\phi}{\tan\theta}\bigg) \nonumber\,, \\ 
	I_{\rm 011} &= I_{\rm source}  \bigg(1-\frac{dz}{dx}\frac{\cos\phi}{\tan\theta}\bigg) \bigg(\frac{dz}{dy}\frac{\sin\phi}{\tan\theta}\bigg) \nonumber\,, \\
\intertext{and}
	I_{\rm 111 } &= I_{\rm source}  \bigg(\frac{dz}{dx}\frac{\cos\phi}{\tan\theta}\bigg) \bigg(\frac{dz}{dy}\frac{\sin\phi}{\tan\theta}\bigg) \nonumber\,. \\
\end{align}

Repeating the interpolation at each subsequent plane (and thus compounding the interpolation error) yields the emergent intensity for point $I_{\rm n_xn_yn_z}$:

\begin{align}\label{eqna2}
	I_{\rm n_x n_y n_z} &= I_{\rm source} \times I_{\rm n_x n_z} \times I_{\rm n_y n_z} \nonumber\,, \\
\intertext{with}
    I_{\rm n_x n_z} &=  \frac{n_z!}{(n_x)!(n_z-n_x)!}\bigg(1- \frac{dz}{dx}\frac{\cos\phi}{\tan\theta} \bigg)^{n_z-n_x} \bigg(\frac{dz}{dx} \frac{\cos\phi}{\tan\theta}\bigg)^{n_x} \nonumber \\
\intertext{and}
    I_{\rm n_y n_z} &= \frac{n_z!}{(n_y)!(n_z-n_y)!}\bigg(1- \frac{dz}{dy}\frac{\sin\phi}{\tan\theta} \bigg)^{n_z-n_y} \bigg(\frac{dz}{dy} \frac{\sin\phi}{\tan\theta}\bigg)^{n_y} \nonumber\,. \\
\end{align}
$I_{\rm n_{x} n_z}$ and $I_{\rm n_{y} n_z}$ are binomial distributions with variances
\begin{align}\label{eqna3}
	\sigma_x &= \sqrt{\ nz\ \frac{dz}{dx} \frac{\cos\phi}{\tan\theta}\ \bigg( 1- \frac{dz}{dx}\frac{\cos\phi}{\tan\theta} \bigg)} \nonumber \\
\intertext{and}	
 \sigma_y &=  \sqrt{\ nz\ \frac{dz}{dy} \frac{\sin\phi}{\tan\theta}\ \bigg( 1- \frac{dz}{dy}\frac{\sin\phi}{\tan\theta}\bigg )} \,.\nonumber \\
 \end{align}
For interpolation in horizontal planes the two-dimension point-spread-function (Eq.~\ref{eqna2}) has $x$ and $y$ widths (Eq.~\ref{eqna3}) that are independent functions of the propagation angle and the number of vertical planes $n_z$ through which the beam has propagated. 

Because the spread in $x$ and $y$ are independent, it is straightforward to show that source intensity is conserved with height.  At any given height $n_z = N_z$ the total intensity
\begin{align}\label{eqna4}
	I_{\rm total} &=  \sum_{n_y = 0}^{N_z} \sum_{n_x = 0}^{N_z} I_{\rm n_x n_y N_{\rm z}} \nonumber \\ 
    		   &=  I_{\rm source} \  \bigg(1- \frac{dz}{dx}\frac{\cos\phi}{\tan\theta} \bigg)^{N_z} \  \bigg(1- \frac{dz}{dy}\frac{\sin\phi}{\tan\theta} \bigg)^{N_z}  \left[\sum_{n_x=0}^{N_z} \frac{N_z!}{(n_x)!(N_z-n_x)!}\bigg( \frac{\frac{dz}{dx}\frac{\cos\phi}{\tan\theta}}{1- \frac{dz}{dx}\frac{\cos\phi}{\tan\theta}}\bigg)^{n_x}\right] \left[ \sum_{n_y=0}^{N_z} \frac{N_z!}{(n_y)!(N_z-n_y)!}\bigg( \frac{\frac{dz}{dy}\frac{\sin\phi}{\tan\theta}}{1- \frac{dz}{dy}\frac{\sin\phi}{\tan\theta}}\bigg)^{n_y}\right]\nonumber \\ 
			   &= I_{\rm source}
\end{align}
where we have used the binomial series expansion $(1 + x)^\alpha = \sum_{k=0}^{\alpha} {{\alpha}\choose{k}} x^k$ with $\alpha=N_z$.

\begin{figure*}[t!]
\centering
\resizebox{\hsize}{!}{
	\subfloat{\includegraphics[width=65mm,height=60mm, trim=0mm 0mm 0mm 0mm, clip=true]{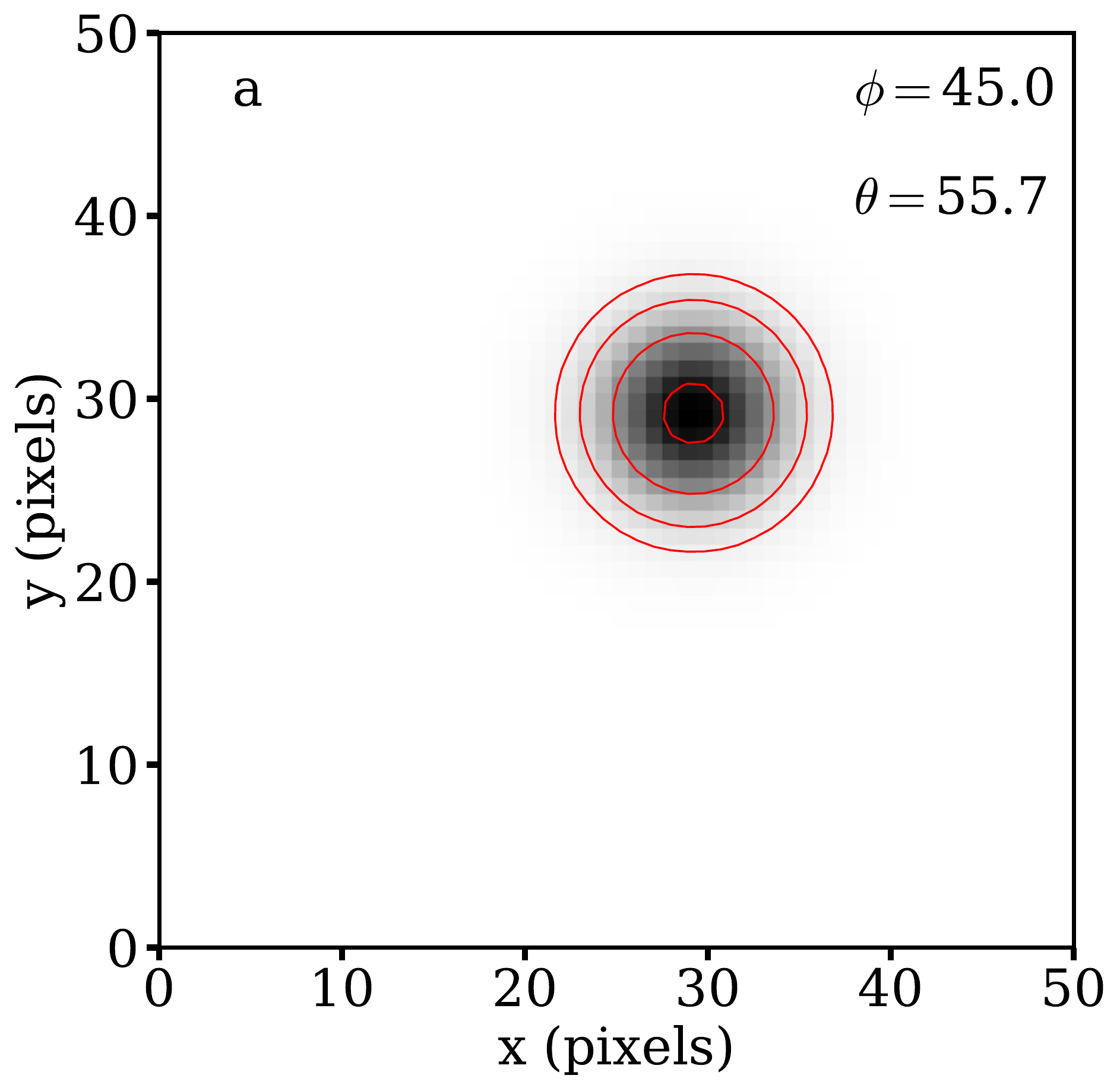}}
	\subfloat{\includegraphics[width=65mm,height=60mm, trim=0mm 0mm 0mm 0mm, clip=true]{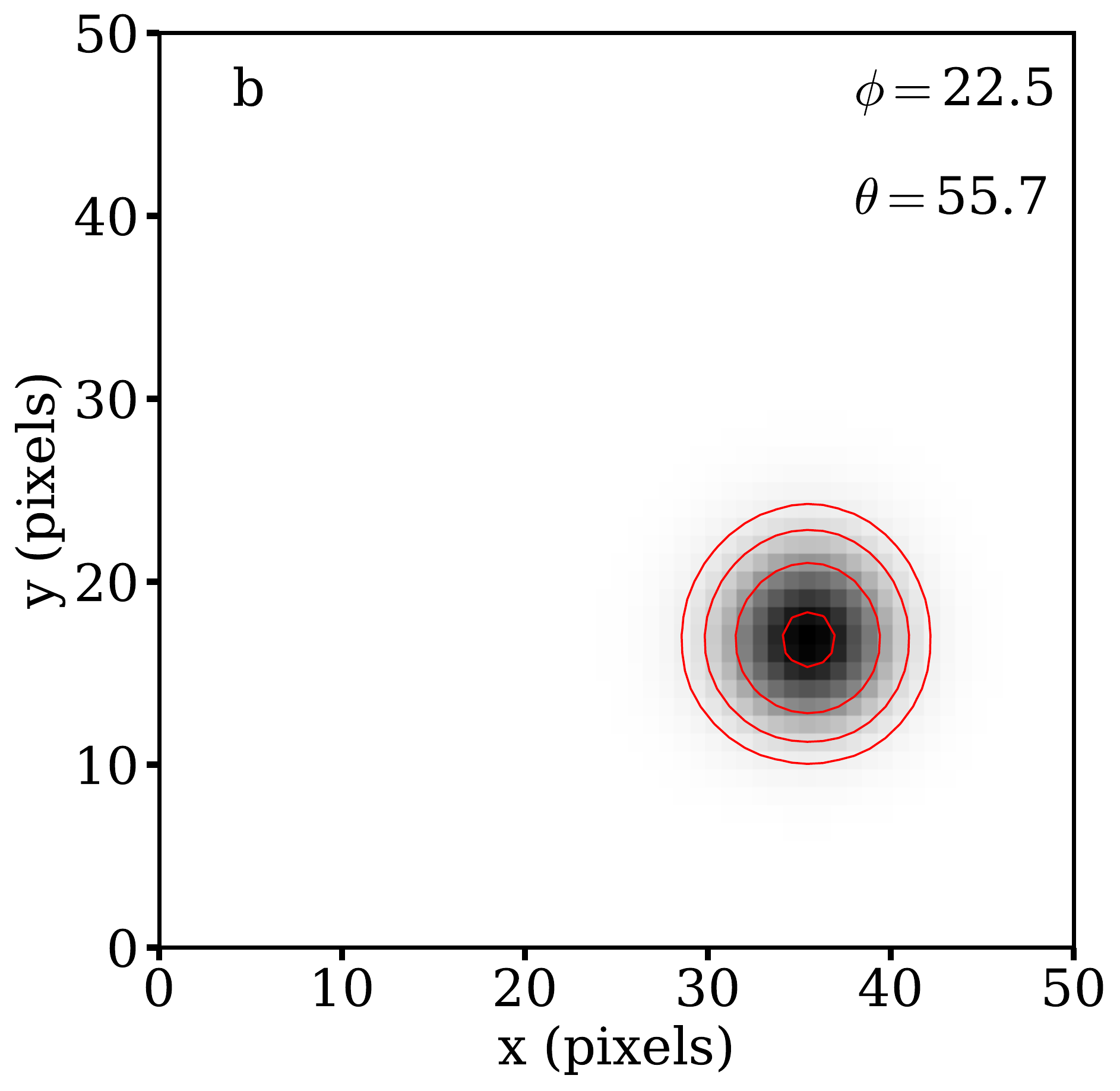}}	
	\subfloat{\includegraphics[width=65mm,height=60mm, trim=0mm 0mm 0mm 0mm, clip=true]{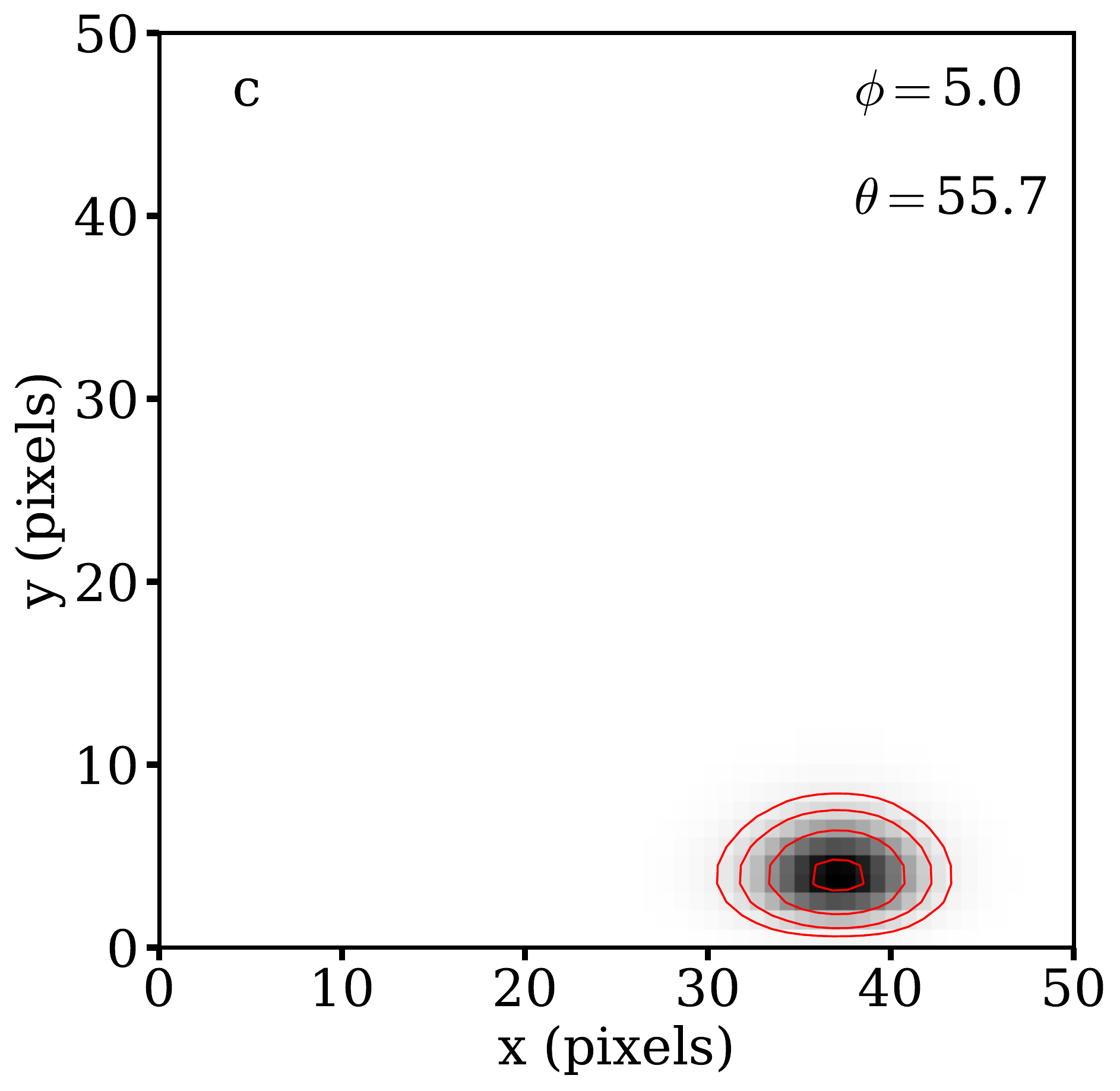}}}
	\caption{PSFs derived from Eq.~\ref{eqna2}. Note that the position of the PSF at the top of the domain simply reflects the propagation angle. ({\it a}) Centered interpolation in the {\it xy} grid. The PSF is circular with $\sigma_x = \sigma_y$. ({\it b} and {\it c}) Non-centered interpolation in the {\it xy} grid, resulting in increasing oblate PSF with decreasing $\phi$. Contours denote the reduction in peak amplitude of 10\%, 50\%, 75\% and 87.5\%. } 
	\label{xyplane_psf}
\end{figure*}

The point-spread-functions (PSF) for three viewing angles are shown in Fig.~\ref{xyplane_psf} and have been scaled to account for apparent foreshortening. Note how the spread in the distributions in the x and y directions are independent.  

\subsection{Solution for ray directions for which interpolation occurs on vertical planes} 
Using the same method as in the previous subsection, we derive the beam diffusion profile when the incident angle is such that the  interpolation occurs in the {\it xz} or {\it yz} plane, as indicated by the right hand panel of Fig.~\ref{SC_diffusion}.  In most short characteristic schemes, the first grid point value is solved employing a long characeristic ray, and we assume the  that is true here for the first upwind grid point to avoid complication and without loss of exactness. 
\begin{figure*}[t!]
\centering
\resizebox{\hsize}{!}{
	\subfloat{\includegraphics[width=65mm,height=60mm, trim=0mm 0mm 0mm 0mm, clip=true]{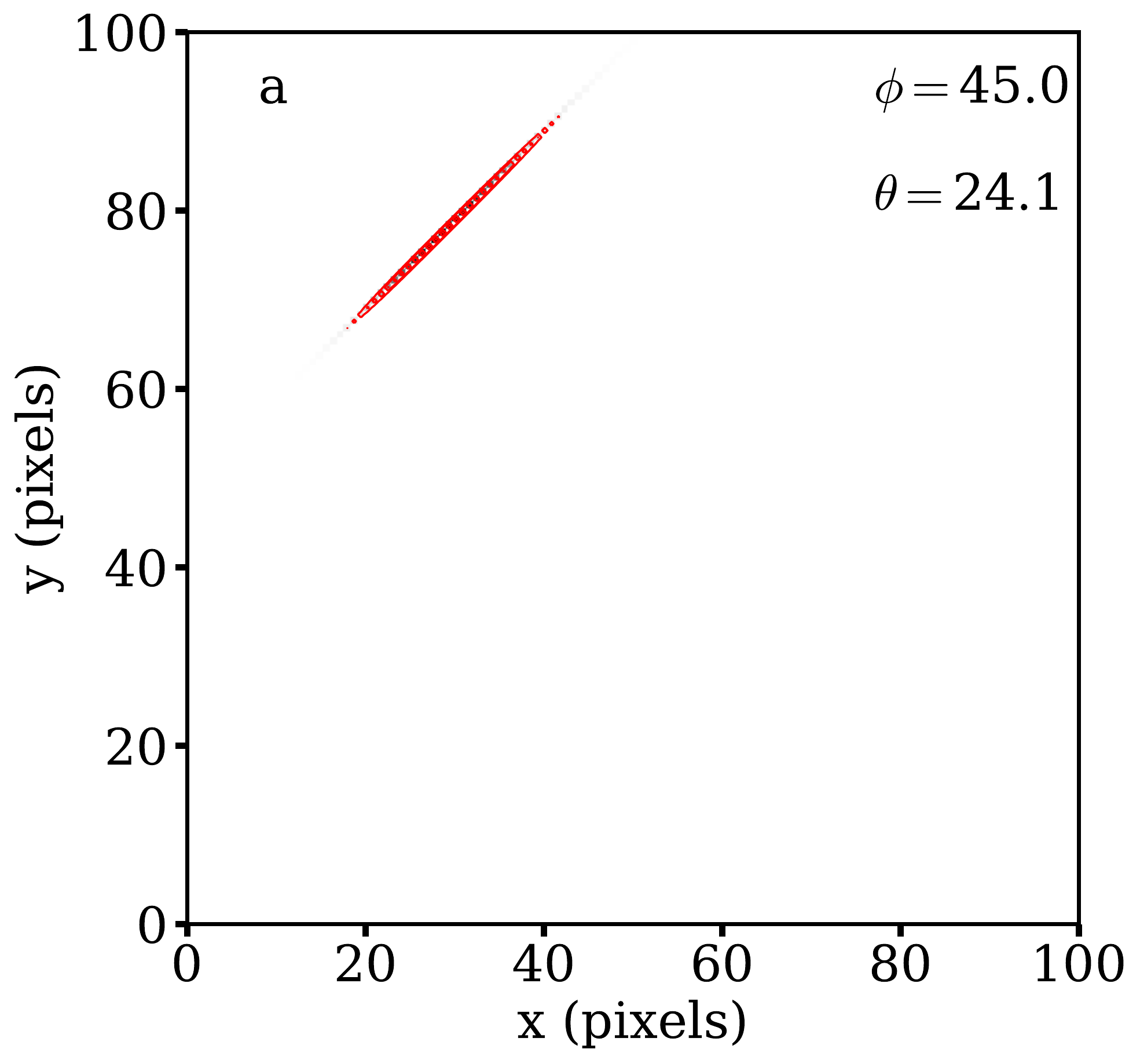}}
	\subfloat{\includegraphics[width=65mm,height=60mm, trim=0mm 0mm 0mm 0mm, clip=true]{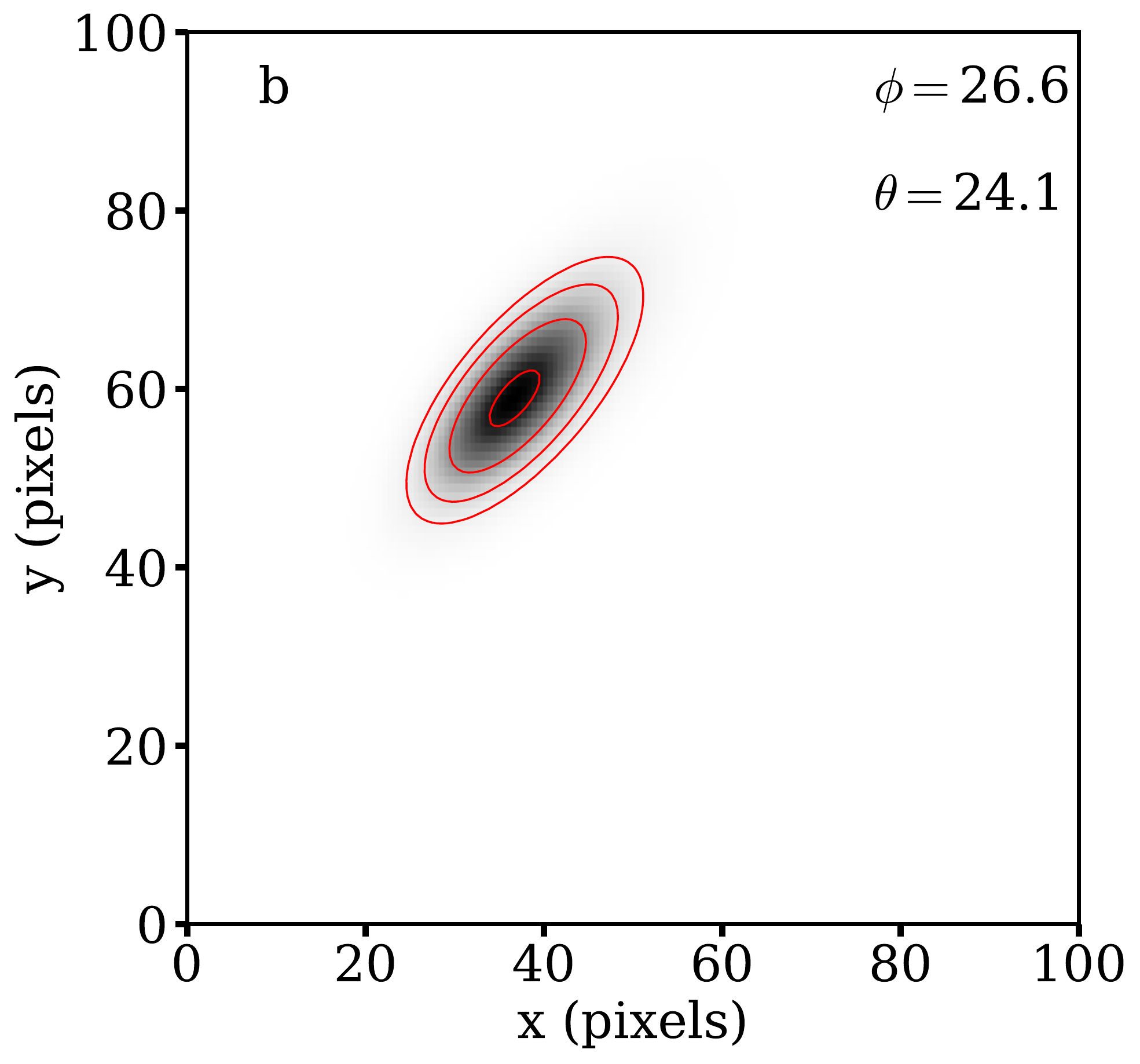}}	
	\subfloat{\includegraphics[width=65mm,height=60mm, trim=0mm 0mm 0mm 0mm, clip=true]{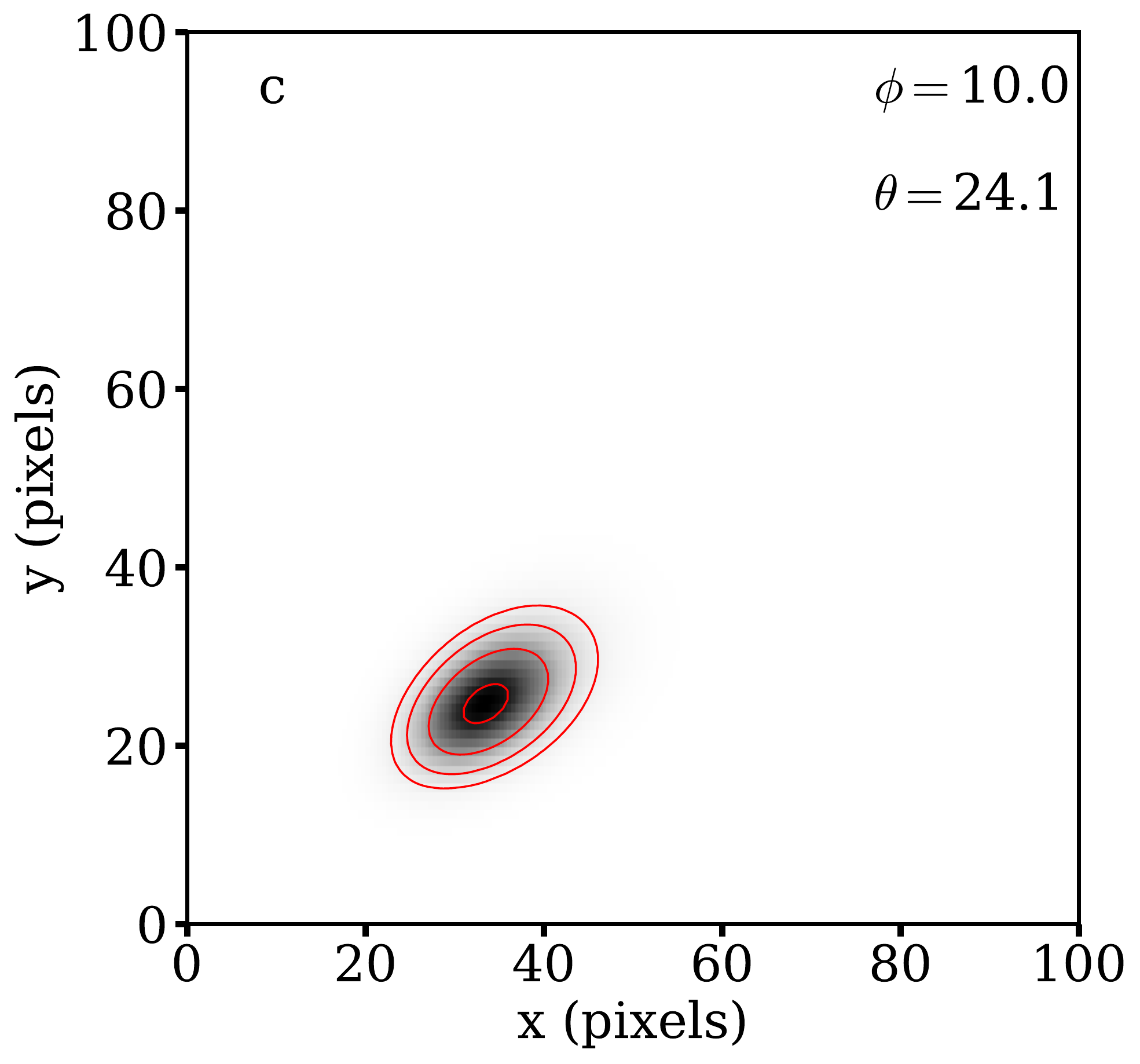}}}
	\caption{PSFs derived from Eq.~\ref{eqna6}. ({\it a}) Interpolation on the z-1 axis due to $\phi = 45$, resulting in smearing only in the inclination direction. ({\it b}) Centered interpolation in the yz grid. Note that $\sigma_x = \sigma_y$ despite the principal axes of the distribution no longer being aligned with x and y as in Fig.~\ref{xyplane_psf}. ({\it c}) Non-centered interpolation in the yz grid which causes an asymmetric distribution. Contours denote the reduction in peak amplitude of 10\%, 50\%, 75\% and 87.5\%.}
	\label{xzyzplane_psf}
\end{figure*}
Unlike for interpolation on horizontal planes, for which only four grid sites on the first horizontal plane downwind of $I_{\rm source}$ are illuminated, interpolation on vertical planes populates the entire horizontal plane down wind of the source.  This is because each subsequent downwind interpolation as one sweep left to right relies on the previous grid point solution.  Thus sweeping from left to right, each successive row front to back in turn, from point $I_{\rm source}$ yields the following: 
\begin{align}\label{eqna5}
	I_{\rm 101} &= I_{\rm source} \bigg(1-\frac{dx}{dy}\frac{\sin\phi}{\cos\phi}\bigg) \bigg(\frac{dx}{dz}\frac{\tan\theta}{\cos\phi}\bigg) \nonumber\,, \\ 
	I_{\rm 201} &= I_{\rm 101}\bigg(1-\frac{dx}{dy}\frac{\sin\phi}{\cos\phi}\bigg)\bigg(1-\frac{dx}{dz}\frac{\tan\theta}{\cos\phi}\bigg)= I_{\rm source} \bigg(1-\frac{dx}{dy}\frac{\sin\phi}{\cos\phi}\bigg)^2 \bigg(1-\frac{dx}{dz}\frac{\tan\theta}{\cos\phi}\bigg) \bigg(\frac{dx}{dz}\frac{\tan\theta}{\cos\phi}\bigg) \nonumber \,,\\ 
    I_{\rm 301} &=I_{\rm 201}\bigg(1-\frac{dx}{dy}\frac{\sin\phi}{\cos\phi}\bigg)\bigg(1-\frac{dx}{dz}\frac{\tan\theta}{\cos\phi}\bigg)= I_{\rm source}  \bigg(1-\frac{dx}{dy}\frac{\sin\phi}{\cos\phi}\bigg)^3\bigg(1- \frac{dx}{dz}\frac{\tan\theta}{\cos\phi}\bigg)^2\bigg(\frac{dx}{dz}\frac{\tan\theta}{\cos\phi}\bigg)  \nonumber \,,\\ 
    I_{\rm 401} &= ... \nonumber\,, \\
	I_{\rm 110} &= 0 \nonumber\,, \\ 
    I_{\rm 111} &= I_{\rm source}  \bigg(\frac{dx}{dy}\frac{\sin\phi}{\cos\phi}\bigg)\bigg(\frac{dx}{dz}\frac{\tan\theta}{\cos\phi}\bigg) \nonumber\,, \\
     I_{\rm 211} &= 2I_{\rm source} \bigg(1-\frac{dx}{dy}\frac{\sin\phi}{\cos\phi}\bigg) \bigg(\frac{dx}{dy}\frac{\sin\phi}{\cos\phi}\bigg) \bigg(1-\frac{dx}{dz}\frac{\tan\theta}{\cos\phi}\bigg) \bigg(\frac{dx}{dz}\frac{\tan\theta}{\cos\phi}\bigg)  \nonumber\,, \\ 
     \intertext{and}
	I_{\rm 311} &= ... \nonumber\,. \\
\end{align}
Continuing this procedure and recognizing the negative binomial distribution with $n_x > n_z$ yields:
 \begin{align}\label{eqna6}
    I_{\rm n_x n_y n_z} &=  I_{\rm source} \times I_{\rm n_x n_z} \times I_{\rm n_y n_x} \nonumber\,, \\
\intertext{with}
    I_{\rm n_x n_z} &= \frac{n_x-1!}{(n_x-n_z)!(n_z-1)!} \bigg(1-\frac{dx}{dz}\frac{\tan\theta}{\cos\phi}\bigg)^{n_x - n_z} \bigg(\frac{dx}{dz}\frac{\tan\theta}{\cos\phi}\bigg)^{n_z} \nonumber \\
\intertext{and}
    I_{\rm n_y n_x} &= \frac{n_x!}{n_y!(n_x-n_y)!} \bigg(1-\frac{dx}{dy}\frac{\sin\phi}{\cos\phi}\bigg)^{n_x - n_y} \bigg(\frac{dx}{dy}\frac{\sin\phi}{\cos\phi}\bigg)^{n_y} \nonumber\,. \\
\end {align}
Note that $I_{\rm n_x n_z}$ in this solution is a negative binomial distribution, while the $I_{\rm n_y n_x}$ is a regular binomial distribution.  $I_{\rm n_x n_z}$, as previously for interpolation on horizontal planes, describes diffusion in $n_x$ and depends on the number of $n_z$ levels through which the ray has passed. $I_{\rm n_y n_x}$ on the other hand 
depends on $n_x$ and $n_y$ as diffusion in the $x$ and $y$ directions are no longer independent. 
The point-spread-functions derived from Eq.~\ref{eqna6} are shown in Fig.~\ref{xzyzplane_psf} for three inclination angles.  They have been scaled to account for apparent foreshortening to show appearance of the PSF at inclined viewing angles. Note that the distributions are no longer aligned with the $x$ and $y$ axes. This is a consequence of the mixing introduced in by the $I_{\rm n_y n_x}$ term. Since the spread in $I_{\rm n_y n_x}$ depends on the $x$ position, the distribution in Fig.~\ref{xzyzplane_psf}$c$ is asymmetric, with larger broadening for larger $x$ values.
The variances associated with these distributions are
\begin{align}\label{eqna7}
	\sigma_x &= \sqrt{n_z \frac{dz}{dx} \frac{\cos\phi}{\tan\theta} \bigg( \frac{dz}{dx}\frac{\cos\phi}{\tan\theta} -1 \bigg)} \nonumber\,, \\
	\intertext{and}
  \sigma_y &=  \sqrt{n_x \frac{dx}{dy} \frac{\sin\phi}{\cos\phi} \bigg( 1- \frac{dx}{dy}\frac{\sin\phi}{\cos\phi}\bigg )} \nonumber\,, \\
 \end{align}
with the variance in the $y$ direction reflecting the asymmetry. 
 
The derivation above is valid for ray angles for which the interpolation occurs in the $yz$ plane.  For angles which the interpolation occurs in the $xz$ plane, the solution is given by Eqs.~\ref{eqna6} and \ref{eqna7} with $x$ and $y$ and $\sin\phi$ and $\cos\phi$ interchanged.

As in \S A1, the specific intensity is conserved. Explicitly,
\begin{align}
I_{\rm total} &=  \sum_{n_x = N_z}^{\infty} \sum_{n_y = 0}^{n_x} I_{\rm n_x n_y N_{\rm z}} \nonumber \\ 
    		   &=  I_{\rm source} \ \bigg(\frac{dx}{dz}\frac{\tan\theta}{\cos\phi}\bigg)^{N_z} \left[\sum_{n_x=N_z}^{\infty} \frac{n_x-1!}{(n_x-N_z)!(N_z-1)!} \bigg(1-\frac{dx}{dz}\frac{\tan\theta}{\cos\phi}\bigg)^{n_x - N_z} \bigg(1-\frac{dx}{dy}\frac{\sin\phi}{\cos\phi}\bigg)^{n_x} \left[ \sum_{n_y=0}^{n_x} \frac{n_x!}{n_y!(n_x-n_y)!}\bigg(\frac{\frac{dx}{dy}\frac{\sin\phi}{\cos\phi}}{1-\frac{dx}{dy}\frac{\sin\phi}{\cos\phi}}\bigg)^{n_y} \right] \right]\nonumber \\
		&= I_{\rm source} \ \bigg(\frac{dx}{dz}\frac{\tan\theta}{\cos\phi}\bigg)^{N_z} \left[\sum_{n_x=N_z}^{\infty} \frac{n_x-1!}{(n_x-N_z)!(N_z-1)!} \bigg(1-\frac{dx}{dz}\frac{\tan\theta}{\cos\phi}\bigg)^{n_x - N_z} \right] \nonumber \\
		 &= I_{\rm source}\, ,
\end{align}
where we have used the binomial and negative binomial series expansions $(1 + x)^\alpha = \sum_{k=0}^{\alpha} {{\alpha}\choose{k}} x^k$ with $\alpha=n_y$ and $1/(1-x)^\beta = \sum_{n=0}^{\infty}{{n+\beta}\choose{n}} x^n$ with $\beta=N_z -1$ and $n = n_x -N_z$.

\bibliographystyle{aasjournal.bst}

\begin{thebibliography}{}

\bibitem[{{Auer} \& {Paletou}(1994)}]{auer1994}
{Auer}, L.~H., \& {Paletou}, F. 1994, \aap, 285, 675

\bibitem[{Bruls {et~al.}(1999)Bruls, Vollm??ller, \& Sch?ºssler}]{bruls_1999}
Bruls, J. H. M.~J., Vollm??ller, P., \& Sch?ºssler, M. 1999, \aap, 348, 233

\bibitem[{{Carlson}(1963)}]{carlson1963}
{Carlson}, B. 1963, in Methods in Computational Physics, Vol. 1, ed. B.~{Alder}
  \& S.~{Fernbach}, 1--42

\bibitem[{{Carlson}(1970)}]{carlson1970}
{Carlson}, B. 1970, {Transport Theory: Discrete Ordinates Quadrature Over the
  Unit Sphere}, Tech. Rep. LA-4554, Los Alamos National Laboratory, Los Alamos,
  NM

\bibitem[{{Carlsson}(2008)}]{carlsson2008}
{Carlsson}, M. 2008, \physscr, T133, 014012

\bibitem[{Collados {et~al.}(2013)Collados, Bettonvil, Cavaller, Ermolli, Gelly,
  P?©rez, Socas-Navarro, Soltau, Volkmer, \& {EST
  Team}}]{collados_european_2013}
Collados, M., Bettonvil, F., Cavaller, L., {et~al.} 2013, \memsai, 84, 379

\bibitem[{Criscuoli(2007)}]{criscuoli_serena_2007}
Criscuoli, S. 2007, PhD thesis, University of Rome Tor Vergata,
  doi:10.5281/zenodo.845496

\bibitem[{{Criscuoli} \& {Rast}(2009)}]{criscuoli2009}
{Criscuoli}, S., \& {Rast}, M.~P. 2009, \aap, 495, 621

\bibitem[{{Davis} {et~al.}(2012){Davis}, {Stone}, \& {Jiang}}]{davis2012}
{Davis}, S.~W., {Stone}, J.~M., \& {Jiang}, Y.-F. 2012, \apjs, 199, 9

\bibitem[{Elmore {et~al.}(2014)Elmore, Rimmele, Casini, Hegwer, Kuhn, Lin,
  McMullin, Reardon, Schmidt, Tritschler, \& W??ger}]{elmore_daniel_2014}
Elmore, D.~F., Rimmele, T., Casini, R., {et~al.} 2014, in , 914707--914707--7

\bibitem[{{Finlator} {et~al.}(2009){Finlator}, {{\"O}zel}, \&
  {Dav{\'e}}}]{finlator2009}
{Finlator}, K., {{\"O}zel}, F., \& {Dav{\'e}}, R. 2009, \mnras, 393, 1090

\bibitem[{Freytag {et~al.}(2002)Freytag, Steffen, \& Dorch}]{freytag2002}
Freytag, B., Steffen, M., \& Dorch, B. 2002, Astronomische Nachrichten, 323,
  213

\bibitem[{Fritsch \& Carlson(1980)}]{fritsch_1980}
Fritsch, F., \& Carlson, R. 1980, SIAM Journal on Numerical Analysis, 17, 238

\bibitem[{Galsgaard \& Nordlund(1996)}]{galsgaard1996}
Galsgaard, K., \& Nordlund, ?. 1996, \jgr, 101, 13445

\bibitem[{{Gudiksen} {et~al.}(2011){Gudiksen}, {Carlsson}, {Hansteen}, {Hayek},
  {Leenaarts}, \& {Mart{\'{\i}}nez-Sykora}}]{gudiksen2011}
{Gudiksen}, B.~V., {Carlsson}, M., {Hansteen}, V.~H., {et~al.} 2011, \aap, 531,
  A154

\bibitem[{{Hayek} {et~al.}(2010){Hayek}, {Asplund}, {Carlsson}, {Trampedach},
  {Collet}, {Gudiksen}, {Hansteen}, \& {Leenaarts}}]{hayek2010}
{Hayek}, W., {Asplund}, M., {Carlsson}, M., {et~al.} 2010, \aap, 517, A49

\bibitem[{{Ibgui} {et~al.}(2013){Ibgui}, {Hubeny}, {Lanz}, \&
  {Stehl{\'e}}}]{ibgui2013}
{Ibgui}, L., {Hubeny}, I., {Lanz}, T., \& {Stehl{\'e}}, C. 2013, \aap, 549,
  A126

\bibitem[{{Kunasz} \& {Auer}(1988)}]{kunasz1988}
{Kunasz}, P., \& {Auer}, L.~H. 1988, \jqsrt, 39, 67

\bibitem[{{Lathrop} \& {Carlson}(1965)}]{lathrop1965}
{Lathrop}, K., \& {Carlson}, B. 1965, {Discrete Ordinates Angular Quadrature of
  the Neutron Transport Equation}, Tech. Rep. LA-3186, Los Alamos National
  Laboratory, Los Alamos, NM

\bibitem[{{Leenaarts} \& {Carlsson}(2009)}]{leenaarts2009}
{Leenaarts}, J., \& {Carlsson}, M. 2009, in Astronomical Society of the Pacific
  Conference Series, Vol. 415, The Second Hinode Science Meeting: Beyond
  Discovery-Toward Understanding, ed. B.~{Lites}, M.~{Cheung}, T.~{Magara},
  J.~{Mariska}, \& K.~{Reeves}, 87

\bibitem[{{Matthews} {et~al.}(2016){Matthews}, {Collados}, {Mathioudakis}, \&
  {Erdelyi}}]{matthews_european_2016}
{Matthews}, S.~A., {Collados}, M., {Mathioudakis}, M., \& {Erdelyi}, R. 2016,
  in \procspie, Vol. 9908, Ground-based and Airborne Instrumentation for
  Astronomy VI, 990809

\bibitem[{{Mihalas} {et~al.}(1978){Mihalas}, {Auer}, \&
  {Mihalas}}]{mihalas1978}
{Mihalas}, D., {Auer}, L.~H., \& {Mihalas}, B.~R. 1978, \apj, 220, 1001

\bibitem[{{Mihalas} \& {Mihalas}(1984)}]{mihalas1984}
{Mihalas}, D., \& {Mihalas}, B.~W. 1984, {Foundations of Radiation
  Hydrodynamics}

\bibitem[{{Pereira} \& {Uitenbroek}(2015)}]{pereira2015}
{Pereira}, T.~M.~D., \& {Uitenbroek}, H. 2015, \aap, 574, A3

\bibitem[{{Rempel}(2014)}]{rempel2014}
{Rempel}, M. 2014, \apj, 789, 132

\bibitem[{Rijkhorst {et~al.}(2006)Rijkhorst, Plewa, Dubey, \&
  Mellema}]{Rijkhorst_hybrid_2006}
Rijkhorst, E.-J., Plewa, T., Dubey, A., \& Mellema, G. 2006, \aap, 452, 907

\bibitem[{Tritschler {et~al.}(2016)Tritschler, Rimmele, Berukoff, Casini, Kuhn,
  Lin, Rast, McMullin, Schmidt, W??ger, \& {DKIST
  Team}}]{tritschler_daniel_2016}
Tritschler, A., Rimmele, T.~R., Berukoff, S., {et~al.} 2016, Astronomische
  Nachrichten, 337, 1064

\bibitem[{{Uitenbroek}(2001)}]{uitenbroek2001}
{Uitenbroek}, H. 2001, \apj, 557, 389

\bibitem[{{{\v S}t{\v e}p{\'a}n} \& {Trujillo Bueno}(2013)}]{stepan2013}
{{\v S}t{\v e}p{\'a}n}, J., \& {Trujillo Bueno}, J. 2013, \aap, 557, A143

\bibitem[{{V{\"o}gler} {et~al.}(2005){V{\"o}gler}, {Shelyag}, {Sch{\"u}ssler},
  {Cattaneo}, {Emonet}, \& {Linde}}]{vogler2005}
{V{\"o}gler}, A., {Shelyag}, S., {Sch{\"u}ssler}, M., {et~al.} 2005, \aap, 429,
  335

\bibitem[{Wray {et~al.}(2015)Wray, Bensassi, Kitiashvili, Mansour, \&
  Kosovichev}]{wray_2015}
Wray, A.~A., Bensassi, K., Kitiashvili, I.~N., Mansour, N.~N., \& Kosovichev,
  A.~G. 2015, arXiv:1507.07999 [astro-ph], arXiv: 1507.07999

\bibitem[{{Zhu} {et~al.}(2015){Zhu}, {Narayan}, {Sadowski}, \&
  {Psaltis}}]{zhu2015}
{Zhu}, Y., {Narayan}, R., {Sadowski}, A., \& {Psaltis}, D. 2015, \mnras, 451,
  1661

\end{thebibliography}

\end{document}